# The role of pseudo-hypersurfaces in non-holonomic motion

D. H. Delphenich ([†])
Spring Valley, OH USA 45370
___

**Abstract:** The geometry of hypersurfaces is generalized to pseudo-hypersurfaces, which are defined by Pfaff equations. The general methods are then applied to modeling the kinematics of motion constrained by a single linear, non-holonomic constraint. They are then applied to the example of a charge moving in an electromagnetic field, and the Lorentz equation of motion is shown to represent a geodesic that is constrained to lie in a pseudo-hypersurface that is defined by the potential 1-form.

**Contents**

**1. Introduction.** – The geometry of curved surfaces is mainly traceable to the 1807 book *Application de l'analyse a la géométrie* by Gaspard Monge, the 1813 book *Développements de géométrie* by Charles Dupin, and the seminal work *Disquisitiones generales circa superficies curvas* of Carl Friedrich Gauss in 1828. Much of the differential geometry of curved spaces of dimension greater than two came about as a result of attempts to generalize the basic ideas that those illustrious geometers introduced.

Of course, there is rarely just one promising direction of generalization to pursue from any given fundamental definition. One direction of generalization that got somewhat less attention than the others was studied by the abbot Pierre Adolphe Issaly in 1889 [**1**] and amounted to generalizing the notion of a surface to a "pseudo-surface." In a more modern context, that would amount to defining a normal 1-form *N* that is not completely integrable. For a surface, it is sufficient to specify that the 1-form is not closed and does admit an integrating factor, so it cannot be exact, but for higher dimensions, complete integrability is equivalent to the vanishing of the Frobenius 3-form $N \wedge d_\wedge N$. The generalization of the definition of a pseudo-surface to that of a "pseudo-hypersurface" then amounts to a mere extension of the dimension of the manifold on which the normal 1-form *N* is defined.

However, the question of the degree of integrability of the exterior differential system *N* = 0 is somewhat more involved in more than two dimensions. As it turns out, what Issaly was defining in the name of a pseudo-surface was closely related to a parallel line of mathematical research that went back to an equally-seminal Latin paper by Johann Friedrich Pfaff in 1815, whose English translation was essentially "A general method for integrating partial differential equations." Basically, when $\phi$ is a 1-form on a differentiable manifold *M* (i.e., a *Pfaffian* form), the exterior differential system $\phi = 0$ is referred to as the *Pfaff equation*. One first solves it algebraically to obtain a hyperplane $\Sigma_x$ in each tangent space $T_xM$, and then says that a submanifold of *M* is an *integral* submanifold of the Pfaff equation iff the tangent space to the submanifold is a subspace of $\Sigma_x$. The maximum dimension of such an integral submanifold is called the *degree of integrability* of the Pfaff equation. One always has integral curves, but one has complete integrability only when there are integral submanifolds whose dimension is one less than that of *M*.

---

([†]) E-mail: feedback@neo-classical-physics.info. Website: neo-classical-physics.info.

The study of the integrability of the Pfaff equation formed the basis for a good many works by the most distinguished mathematicians of the Nineteenth and early Twentieth Centuries and advanced both analysis and geometry in the process. The names of contributors included the likes of Alfred Clebsch, Georg Frobenius, Gaston Darboux, Élie Cartan, and Edouard Goursat. The contributions to the methodology of solving the Pfaff problem by Simeon Denis Poisson and Carl Gustav Jacob Jacobi led to the work of Sophus Lie on the symmetries of systems of differential equations.

A particularly deep extension of the concept of the Pfaff equation was that of a "Pfaffian system," in which one has more than one Pfaffian form, and they are required to vanish simultaneously. Algebraically, if there are *p* 1-forms that must vanish simultaneously then the hyperplane $\Sigma_x$ will become the intersection of those *p* 1-forms, and will therefore have a dimension of *n* – *p* – 1 if the dimension of *M* is *n*. The definition of an integral submanifold does not change, but the existence of integral submanifolds and the degree of integrability becomes much more involved. The most definitive result was the extension of the Cauchy-Kovalevskaya theorem for partial differential equations to Pfaffian systems that was made by Cartan in 1901 and refined somewhat by Erich Kähler in 1931.

One of the most fruitful applications of the techniques that grew out of the study of the Pfaff equation was to the mechanics of systems that are subject to non-holonomic constraints. The distinction between "holonomic" and "non-holonomic" seems to have originated with Heinrich Hertz in his intriguing, if not somewhat neglected, treatise on the principles of mechanics [**2**], which was published posthumously in the year of his death 1894. Basically, a constraint that is defined by a Pfaffian system on a configuration manifold is holonomic when that system is completely integrable and non-holonomic otherwise.

Some of the more geometric and analytical studies of non-holonomic constraints in the late Nineteenth and early Twentieth Century were made by R. Lipschitz (1874, [**3**]), Alexander Voss (1885, [**4**]), Jacques Hadamard (1895, [**5**]), S. Dautheville (1909, [**6**]), P. Voronetz (1911, [**7**]), Ivan Tzenoff (1920-25, [**8**]), Paul Appel (1899-1925, [**9**]), Georg Hamel (1904-35, [**10**]), and Étienne Delassus (1912-13, [**11**]). (A more recent discussion of non-holonomic motion can be found in the book by Neimark and Fufaev [**12**].)

Once the importance of non-holonomic constraints was established in physics, there were some attempts to essentially generalize Issaly's concept of a pseudo-surface to a "non-holonomic manifold," although not all of the researchers seemed to have been aware of Isally's work. Some of the more notable advances were due to Reinhold von Lilienthal (1888-99, [**13**]), Gheorghe Vranceanu (1926, [**14**]), Zdenek Horak (1927-35, [**15**]), John L. Synge (1927-28, [**16**]), D. Sintzov (1929, [**17**]), Jan Schouten (1928-29, [**18**]), Schouten and E. R. van Kampen (1930, [**19**]), P. Franklin and C. L. E. Moore (1931, [**20**]), A. Wundheiler (1931-32, [**21**]), and V. Wagner (1943, [**22**]), which were typically confined to the Pfaff equation, rather than a Pfaff system. Since integral curves of the Pfaff equation will always exist, a fundamental question was that of how the various distinguished curves on hypersurfaces that relate to the eigenvalues and eigenvectors of the second fundamental form generalize to the non-integrable case. In particular, the question of geodesics attracted a lot of attention, since the equivalence of geodesics and "normal curves" breaks down when one no longer has the complete integrability of the Pfaff equation.

The following treatise is an attempt to specialize the concept of non-holonomic manifold to that of a pseudo-hypersurface, which is then defined to be the codimension-one sub-bundle of the tangent bundle to a differentiable manifold, and then apply it to the study of mechanical systems that are constrained by a single linear non-holonomic constraint. In order to stay as close as possible to the published work on non-holonomic motion that is due to the physicists, we shall restrict our consideration to $\mathbb{R}^{n+1}$ as the basic manifold on which everything else is defined. Although that will ignore certain topological subtleties, nonetheless, it will become clear in what follows that simply defining the basic machinery of pseudo-hypersurfaces and their applications to physics is already an ambitious undertaking. One can safely treat the topological issues as advanced topics in that light.

The particular example of non-holonomic motion that we shall ultimately consider is that of a charge that moves in electromagnetic field. The pseudo-hypersurface that pertains to that example is the horizontal sub-bundle of the $U(1)$-principal bundle that serves as the gauge structure for electromagnetism and is defined by the vanishing of the connection 1-form that one uses for the electromagnetic potential. One then finds that the constrained geodesics of that pseudo-hypersurface satisfy the Lorentz force equation for a suitable choice of Lagrange multiplier.

Hopefully, the table of contents that is listed above will serve as an adequate summary of the sections to follow.

**2. The geometry of hypersurfaces.** – Our treatment of the geometry of hypersurfaces is a generalization of the classical geometry of surfaces (e.g., Eisenhart, [**23**]), but a specialization of the more modern treatment that one might in find in vol. 2 of Kobayashi and Nomizu [**24**]. In particular, since we will be treating the embedding manifold $\mathbb{R}^{n+1}$ as a flat manifold, it will not be necessary to introduce covariant derivatives, and we can work with just partial derivatives.

We shall begin by specifying some things that will be used consistently throughout.

*a. Notations and conventions.* – $\mathbb{R}^{n+1}$ will be given the canonical coordinate system that takes each point $x = (x^1, \ldots, x^{n+1})$ in $\mathbb{R}^{n+1}$ to its coordinates $x^i(x) = x^i$, $i = 1, \ldots, n+1$. Hence, there will also be a natural frame field $\{\partial_i = \partial / \partial x^i, i = 1, \ldots, n+1\}$ on $\mathbb{R}^{n+1}$ and a reciprocal natural coframe field $\{dx^i, i = 1, \ldots, n+1\}$. They are reciprocal in the sense that:

$$dx^i (\partial_j) = \delta^i_j . \tag{2.1}$$

The vector space $\mathbb{R}^{n+1}$ will typically be given the Euclidian metric (or scalar product):

$$\delta = \delta_{ij} \, dx^i \, dx^j, \tag{2.2}$$

which makes the natural frame field orthonormal:

$$\delta(\partial_i, \partial_j) = \delta_{ij} . \tag{2.3}$$

The multiplication of 1-forms in the right-hand side of (2.2) is implicitly the symmetrized tensor product.

In the case of Minkowski space, the hyperbolic normal metric:

$$\eta = \eta_{\mu\nu} \, dx^\mu \, dx^\nu, \qquad \eta_{\mu\nu} = \text{diag}\,[1, -1, -1, -1], \qquad \mu, \nu = 0, 1, 2, 3 \tag{2.4}$$

will be used instead of $\delta$. The orthonormality of the natural frame field will then take the form:

$$\eta(\partial_\mu, \partial_\nu) = \eta_{\mu\nu} . \tag{2.5}$$

There is also a metric that is defined on the dual space $\mathbb{R}^{n+1*}$ to $\mathbb{R}^{n+1}$:

$$\delta = \delta^{ij} \, \partial_i \, \partial_j , \tag{2.6}$$

which makes the natural coframe field orthonormal. The component matrix $\delta^{ij}$ is the inverse to $\delta_{ij}$, so:

$$\delta^{ik} \delta_{kj} = \delta^i_j . \tag{2.7}$$

As a result of the constancy of the component matrices, the Levi-Civita connection for the metrics $\delta$ and $\eta$ will both vanish, so the ordinary differentials and derivatives can be used, rather than the covariant ones. However, in the more general case when the ambient manifold of the embedding or submersion is an $(n+1)$-dimensional differentiable manifold with some sort of connection defined on it, one can simply perform a "minimal coupling" of the connection by replacing ordinary differentials and derivatives with covariant ones. The reason for dealing with the restricted case here is that one can focus on the geometric objects that come about solely as a result of the constraints on the motion and have no origin in the complexities of the ambient space, such as ones that might be due to strong gravitational fields.

*b. Defining hypersurfaces.* –There are basically two ways of defining a hypersurface in $\mathbb{R}^{n+1}$:

1. As a *locus*: In this case, one embeds an *n*-dimensional (parameter) manifold $S$ as a submanifold by way of a differentiable map $x : S \to \mathbb{R}^{n+1}$, $u^a \mapsto x^i(u)$. As an embedding, the map $x$ is injective, so its differential map $dx|_x : T_x S \to \mathbb{R}^{n+1}$, $\mathbf{v} \mapsto dx|_x(\mathbf{v})$ will also be injective, and one specifies that the image of $x$ does not intersect itself, either.

If $\mathbf{v} = v^a \partial_a$ then the components of the vector $dx|_x(\mathbf{v})$ with respect to the natural frame field on $\mathbb{R}^{n+1}$ will be:

$$[dx|_x(\mathbf{v})]^i = \frac{\partial x^i}{\partial u^a}(u) v^a. \tag{2.8}$$

We shall also use the abbreviation:

$$x^i_{,a} = \frac{\partial x^i}{\partial u^a}. \tag{2.9}$$

2. As an *envelope*: In this case, one has a smooth function $\phi : \mathbb{R}^{n+1} \to \mathbb{R}$, $x \mapsto \phi(x)$, and specifies that a hypersurface is a level set of $\phi$; that is, the set of all points $x \in \mathbb{R}^{n+1}$ such that $\phi(x)$ has the same value for all of them.

In this way of defining a hypersurface, there is a natural 1-form $N$ that is defined by $\phi$, namely:

$$N = d\phi = \phi_{,i}\, dx^i, \tag{2.10}$$

which will be assumed to be non-zero at every point. (Hence, the function $\phi$ will represent a *submersion* of $\mathbb{R}^{n+1}$ in $\mathbb{R}$.)

A basic property of a 1-form that is defined in this way (viz., an *exact* 1-form) is that its *exterior derivative* must vanish identically ([1]):

$$0 = d_\wedge N = \tfrac{1}{2}(\partial_i N_j - \partial_j N_i)\, dx^i \wedge dx^j. \tag{2.11}$$

When one substitutes $\partial_i \phi$ for $N_i$, this will follow identically from the equality of mixed partial derivatives for continuously twice-differentiable functions.

The 1-form $N$ can be characterized as the normal 1-form to the hypersurfaces thus-defined, since it defines a hyperplane $\Sigma_x$ in the tangent space $T_x$ at each $x \in \mathbb{R}^{n+1}$ by its vanishing. Hence, a vector $\mathbf{v} \in T_x$ will belong to $\Sigma_x$ iff:

$$0 = N(\mathbf{v}) = \phi_{,i}\, v^i. \tag{2.12}$$

One can also denote this situation by the exterior differential system:

---

[1] One then says that the 1-form $N$ is *closed*.

$$N = 0, \tag{2.13}$$

which is also a Pfaff equation ([1]).

Due to (2.12), one can then regard $\Sigma_x$ as the tangent hyperplane to the level hypersurface at $x$. If one raises the index on $\phi_{,i}$ using the metric $\delta$ then one will get a vector field:

$$\mathbf{N} = N^i \partial_i, \qquad N^i \equiv \delta^{ij} \phi_{,j}, \tag{2.14}$$

that is basically the gradient of $\phi$, and it has the property that it is orthogonal to every vector $\mathbf{v}$ in $\Sigma_x$ since:

$$0 = N(\mathbf{v}) = \phi_{,i} v^i = \delta_{ij} N^i v^j = \delta(\mathbf{N}, \mathbf{v}). \tag{2.15}$$

Hence, one can regard $\mathbf{N}$ as a normal vector field to the hyperplane field $\Sigma$, which associates each $x \in \mathbb{R}^{n+1}$ with the hyperplane $\Sigma_x$.

Since the vector field $\mathbf{N}$ is everywhere-nonzero, it can be normalized to a unit vector field $\hat{\mathbf{N}}(s)$ by multiplying each vector $\mathbf{N}(x)$ by $\|\mathbf{N}\|^{-1}$. Hence:

$$\mathbf{N} = \|\mathbf{N}\| \hat{\mathbf{N}}(s). \tag{2.16}$$

That will not change the nature of its orthogonal complement. Similarly, normalizing $d\phi$ by multiplying it by:

$$\alpha = \|d\phi\|^{-1} = \|\mathbf{N}\|$$

will not change its annihilating subspaces, but it will change the nature of its differential, since:

$$d(\alpha \, d\phi) = d\alpha \otimes d\phi + \alpha \, d^2\phi.$$

However, when this tensor is applied to two vectors $\mathbf{v}$, $\mathbf{w}$ in the annihilating plane of $d\phi$, the first term will vanish:

$$d(\alpha \, d\phi)(\mathbf{v}, \mathbf{w}) = d\alpha(\mathbf{v}) \, d\phi(\mathbf{w}) + \alpha \, d^2\phi(\mathbf{v}, \mathbf{w}) = \alpha \, d^2\phi(\mathbf{v}, \mathbf{w}).$$

Hence, as long as we are dealing with vectors in $\Sigma_x$, there is no loss in generality in restricting $d\phi$ to be a unit covector and restricting $\mathbf{N}$ to be a unit vector.

Although $d^2\phi$ is a symmetric, second-rank covariant tensor, $d(\alpha \, d\phi)$ is not generally symmetric, since its antisymmetric part will be $d\alpha \wedge d\phi$.

Actually, since $\phi(x)$ associates every point $x \in \mathbb{R}^{n+1}$ with a real number, and each number in the image of $\phi$ defines a hypersurface in $\mathbb{R}^{n+1}$, there will be a hypersurface through each point in $\mathbb{R}^{n+1}$. Thus, one finds that defining a hypersurface as an envelope actually gives one more than one gets from a locus, which is only a single hypersurface. What one gets from an envelope is an elementary example of a *foliation* of $\mathbb{R}^{n+1}$ that has codimension one; i.e., a decomposition of $\mathbb{R}^{n+1}$ into a disjoint union of "leaves," which take the form of hypersurfaces in the present case.

---

([1]) The author has compiled a monograph [**25**] on the applications of the theory of the Pfaff equation to various branches of physics.

When one starts with a hypersurface as an envelope ($\phi$ = const.), an embedded hypersurface (i.e., locus) $x : S \to \mathbb{R}^{n+1}$ will then take the form of an *integral submanifold* of the Pfaff equation $N = 0$. That is, when one composes the embedding with the function $\phi$, the composed function $\phi \cdot x : S \to \mathbb{R}$, $u \mapsto \phi(x(u))$ will be a constant function. One can also say that the "pull-back" $x^*N$ of the 1-form $N$ on $\mathbb{R}^{n+1}$ by the embedding $x$ (which will then be a 1-form on $S$) must vanish:

$$0 = x^*N = \left(\phi_{,i}\, x^i_{,a}\right) du^a , \qquad (2.17)$$

which can be given the component form:

$$0 = \frac{\partial \phi}{\partial x^i} \frac{\partial x^i}{\partial u^a}. \qquad (2.18)$$

A useful property of pull-back maps is that they commute with the differential map:

$$x^* d\phi = d\,(x^*\phi) \equiv d\,(\phi \cdot x).$$

In the general case, as long as $N$ is not zero anywhere (otherwise, it would be a *singular* Pfaffian form), it will still define an annihilating hyperplane $\Sigma_x$ at every $x \in \mathbb{R}^{n+1}$. However, the existence of integral submanifolds of the Pfaff equation (2.13) would not be guaranteed to the same extent. One defines the *degree of integrability* of the Pfaff equation (2.13) to be the largest dimension of its integral submanifolds. At the very least, integral curves always exist; that is, integral submanifolds of dimension one. A Pfaff equation is *completely integrable* iff the degree of integrability is $n$, which would then define a true hypersurface ([1]).

Actually, $N$ does not have to be exact in order for its Pfaff equation to be completely integrable. It can also include an "integrating factor," such as when $N = \lambda\, d\phi$. That is because as long as $\lambda$ is nowhere-vanishing, the vanishing of $N$ will be equivalent to the vanishing of $d\phi$. Hence, the integral submanifolds will again be level hypersurfaces of $\phi$. In such a case, one finds that:

$$N \wedge d_\wedge N = \lambda\, d\phi \wedge d\lambda \wedge d\phi = 0, \qquad (2.19)$$

---

([1]) For more details on the degree of integrability of a Pfaff equation, see the author's monograph [**25**].

since the exterior product $\wedge$ is antisymmetric. By Frobenius's theorem ([1]), that condition is equivalent to the complete integrability of the Pfaff equation defined by $N$.

*c. Adapted frames and coframes.* – A particularly useful choice of frame $\{\mathbf{e}_i, i = 1, \ldots, n + 1\}$ in a tangent space $T_x$ to a point $x \in \mathbb{R}^{n+1}$ is one that is *adapted* to the direct sum decomposition:

$$T_x = \Sigma_x \oplus [\mathbf{N}]_x,$$

where $[\mathbf{N}]_x$ is the line that is generated by the unit vector $\mathbf{N}$. For such a frame, one of the members (we shall use $\mathbf{e}_{n+1}$) generates $[\mathbf{N}]_x$ (i.e., coincides with $\mathbf{N}$), and the other $n$ of them span the $n$-dimensional subspace $\Sigma_x$. Hence, a vector $\mathbf{v}$ will lie in $\Sigma_x$ iff it can be written in the form:

$$\mathbf{v} = v^a \, \mathbf{e}_a, \qquad a = 1, \ldots, n. \tag{2.20}$$

Since the Pfaff equation $d\phi = 0$ is completely-integrable, one can use the embedding $x : S \to \mathbb{R}^{n+1}$ to define a (typically curvilinear) coordinate system $\{u^a, a = 1, \ldots, n\}$ on its image, so the natural frame field $\{\partial_a, a = 1, \ldots, n\}$ will define an adapted frame on the tangent spaces $\Sigma_x$ to the image. As long as $d\phi$ is everywhere non-zero, one can extend the coordinate system $\{u^a, a = 1, \ldots, n\}$ on the image to a coordinate system $\{u^a, \phi\}$ on $\mathbb{R}^{n+1}$, and the frame field $\{\partial_a, \partial_\phi\}$ will be an adapted frame field with respect to the foliation of $\mathbb{R}^{n+1}$ by level hypersurfaces.

The dual situation applies to the direct-sum decomposition:

$$T_x^* = \Sigma_x^* \oplus [N]_x.$$

That is, a coframe $\{\theta^i, i = 1, \ldots, n + 1\}$ is *adapted* to that decomposition when one of the members (we shall use $\theta^{n+1}$) generates $[N]_x$ and the other $n$ span the complementary subspace $\Sigma_x^*$. Hence, a covector $\alpha$ will belong to $\Sigma_x^*$ iff it can be written in the form:

$$\alpha = \alpha_a \, \theta^a, \qquad a = 1, \ldots, n. \tag{2.21}$$

One easily sees that the reciprocal coframe to an adapted frame will be adapted.

The adapted coordinate system $\{u^a, \phi\}$ will also imply an adapted natural coframe field $\{du^a, d\phi\}$.

*d. The first fundamental form.* – When one has an embedding $x : S \to \mathbb{R}^{n+1}$ of a manifold $S$ in $\mathbb{R}^{n+1}$, one can pull the metric $\delta$ on $\mathbb{R}^{n+1}$ back to a metric $g = x^* \delta$ on $S$. By definition, when $g$ is defined on $S$ in that way, the embedding $x$ will become an *isometric embedding*.

---

([1]) Some good references on Frobenius's theorem are [**26, 27**].

If **v** and **w** are tangent vectors to $S$ at some point $u$ then the value of the pull-back metric $g = x^* \delta$ will be:

$$g(\mathbf{v}, \mathbf{w}) = x^* \delta(\mathbf{v}, \mathbf{w}) = \delta(dx|_u(\mathbf{v}), dx|_u(\mathbf{w})). \qquad (2.22)$$

This means, in particular, that **v** and **w** will be orthogonal with respect to $g$ iff their images $dx|_u(\mathbf{v})$ and $dx|_u(\mathbf{w})$ are orthogonal with respect to $\delta$, and similarly a tangent vector **v** to $S$ will be a unit vector with respect to $g$ iff its image under $x$ is a unit vector for $\delta$.

This metric $g$ is commonly referred to as the *first fundamental form* for the hypersurface that is defined by $x$.

If $x^i$ are coordinates on $\mathbb{R}^{n+1}$ that make the components of $\delta$ equal to $\delta_{ij}$ and $u^a$ are (local) coordinates on $S$ then $g$ will take the form:

$$g = g_{ab}\, du^a\, du^b, \qquad (2.23)$$

with:

$$g_{ab} = \delta_{ij} \frac{\partial x^i}{\partial u^a} \frac{\partial x^j}{\partial u^b}. \qquad (2.24)$$

*e. The second fundamental form.* – The differential of the normal 1-form $dN$:

$$dN = \phi_{,i,j}\, dx^i\, dx^j \qquad (2.25)$$

plays an important role in the geometry of the hypersurface. One refers to the pull-back of $dN$ by the embedding $x: S \to \mathbb{R}^{n+1}$ of a codimension-one submanifold that has $\Sigma$ for its tangent spaces, namely:

$$H = x^* dN = \left( \frac{\partial x^i}{\partial u^a} \frac{\partial x^j}{\partial u^b} \frac{\partial^2 \phi}{\partial x^i \partial x^j} \right) du^a du^b, \qquad (2.26)$$

as the *second fundamental form* for the hypersurface that is defined by $x$. $H$ is then a tensor field on the embedded manifold $S$ whose components with respect to the natural coframe on $S$ are:

$$H_{ab} = \frac{\partial x^i}{\partial u^a} \frac{\partial x^j}{\partial u^b} \frac{\partial^2 \phi}{\partial x^i \partial x^j}. \qquad (2.27)$$

Due to the symmetry of the partial derivatives, $H$ is a symmetric, second-rank covariant tensor field on $S$, and thus defines a symmetric bilinear form on $T(S)$:

$$H(\mathbf{v}, \mathbf{w}) = H_{ab}\, v^a\, w^b. \qquad (2.28)$$

$H$ does not have to be non-degenerate, since it can have zero eigenvalues.

*f. Eigenvalues of H*. – Since $H_{ab}$ is symmetric in $a$ and $b$, the $n \times n$ matrix:

$$H_b^a = g^{ac} H_{cb} \tag{2.29}$$

will have real eigenvalues $\kappa_a$, $a = 1, \ldots, n$ and orthogonal eigenvectors.

Hence, in a principal frame (which is composed of eigenvectors), one will have:

$$H_b^a = \text{diag}\,[\kappa_1, \ldots, \kappa_n]. \tag{2.30}$$

The eigenvalues $\kappa_a$ of $H$ at each point of $S$ are called the *principal curvatures* of the hypersurface $S$. In particular, when one holds $\kappa$ constant, the equation:

$$\kappa = H(\mathbf{t}, \mathbf{t}) = H_{ab}\, t^a t^b, \qquad a, b = 1, \ldots, n \tag{2.31}$$

will define a quadric hypersurface in each tangent space $T_u S$, which will be called a *fundamental quadric for H* at $u$; when $\kappa$ is a principal curvature, it will be called a *principal quadric*. In particular, the quadric that is defined by:

$$H(\mathbf{v}, \mathbf{v}) = 1 \tag{2.32}$$

is referred to as the *Dupin indicatrix*. In the principal frame for $H$, it will take the form:

$$\frac{(X^1)^2}{\rho_1} + \cdots + \frac{(X^n)^2}{\rho_n} = 1 \tag{2.33}$$

for the non-zero eigenvalues. ($\rho_a = 1/\kappa_a$ is the *radius of curvature* that is associated with $\kappa_a$, which can be positive or negative.)

One can then classify the points of $S$ according to the signature type of $H$ as a quadratic form on tangent vectors. For instance:

1. A point is *elliptic* when all $\kappa_a$ are non-zero and have the same sign.

2. A point is *hyperbolic* when all $\kappa_a$ are non-zero and have differing signs. One can then decompose the tangent space $T_u S$ into a direct sum $T_u^- S \oplus T_u^+ S$ of vectors with negative eigenvalues and vectors positive ones, respectively.

3. A point is *parabolic* when some of the $\kappa_a$ are zero.

4. A point is *umbilic* when all $\kappa_a$ are equal. In particular, when they are all zero, the point is flat.

In the general case, the tangent space $T_u S$ can be decomposed into a direct sum:

$$T_u S = T_u^- S \oplus T_u^0 S \oplus T_u^+ S,$$

whose summands correspond to negative, zero, and positive eigenvalues, respectively.

*g. Invariants of H.* – One has two frame-invariants for $H$, namely, the trace of $H$:

$$\text{Tr } H = H_a^a = \sum_{a=1}^{n} \kappa_a, \tag{2.34}$$

which defines the *mean curvature* of the hypersurface at each point by way of:

$$\bar{\kappa} = \frac{1}{n} \text{Tr } H, \tag{2.35}$$

and the determinant:

$$K = \det H = \prod_{a=1}^{n} \kappa_a, \tag{2.36}$$

which defines its *Gaussian curvature*. Our choice of terminology is based upon the fact that when $n = 2$, this will be the usual Gaussian curvature of a surface. Note that the mean curvature at an umbilic will be equal to any of the principal curvatures (call it $\kappa$), and its Gaussian curvature will be $\kappa^n$.

If one goes back to the definition of **N** as:

$$\mathbf{N} = N^i \, \partial_i, \qquad N^i = \delta^{ij} \partial_j \phi,$$

and uses the adapted coordinate system $\{u^a, \phi\}$ then $dx^{n+1} = d\phi = N$ which makes the only non-zero component of $N$ (namely, $N_{n+1}$) equal to 1. Hence:

$$\partial_i N_{n+1} = \partial_i \partial_{n+1} \phi = 0 \text{ for all } i = 1, \ldots, n+1,$$

which will make:

$$\text{div } \mathbf{N} = \delta^{ij} \partial_i \partial_j \phi = \delta^{ab} \partial_a \partial_b \phi, \qquad a, b = 1, \ldots, n.$$

Thus:

$$H_a^a = g^{ab} \frac{\partial x^i}{\partial u^a} \frac{\partial x^j}{\partial u^b} H_{ij} = g^{ab} \frac{\partial x^i}{\partial u^a} \frac{\partial x^j}{\partial u^b} \partial_{ij} \phi = g^{ab} \frac{\partial x^c}{\partial u^a} \frac{\partial x^d}{\partial u^b} \partial_{cd} \phi = \delta^{ab} \partial_{ab} \phi = \text{div } \mathbf{N},$$

which can be written as:

$$n\bar{\kappa} = \text{div } \mathbf{N}. \tag{2.37}$$

Hence, the mean curvature of the hypersurface is due to the divergence of the normal vector field. Since $N$ is exact in the present case, we also have that:

$$n\bar{\kappa} = \Delta\phi. \tag{2.38}$$

If we define a hypersurface to be *minimal* when its mean curvature vanishes then that will imply the:

**Theorem:**

*The hypersurface defined by the level sets of $\phi$ is minimal iff $\phi$ is harmonic.*

Such a hypersurface will then take the form of an equipotential surface.

One sees that there are actually two distinct directions of generalization leading away from the Gaussian curvature of surfaces, one of which is the extension that we just made, and the more traditional direction leads to the Riemann curvature tensor, whose single independent component in the case of surfaces will be proportional to the Gaussian curvature. Interestingly, Kronecker [**28**] had once looked into using essentially *K* (times a volume element) as a way of extending Gauss's *curvatura integra* (viz., integral of the curvature form over the manifold, which is also called the "total curvature" by some) and found that one also got an extension of the Gauss-Bonnet theorem to *n*-dimensional submanifolds for any *n*, as opposed to the extension that Chern made later using the Riemann curvature tensor, which would only work for even-dimensional manifolds, since one would be looking at exterior powers of the curvature 2-form.

*h. Curves in hypersurfaces.* – A curve $x : \mathbb{R} \to \mathbb{R}^{n+1}$, $s \mapsto x(s)$, is said to *lie in a hypersurface* if either:

1. (Locus) There is a curve $u : \mathbb{R} \to S$, $s \mapsto u(s)$ in *S*, and *S* is embedded in $\mathbb{R}^{n+1}$ as a submanifold by a map $x : S \to \mathbb{R}^{n+1}$, $u \mapsto x(u)$, so one can define the curve $x(s)$ to be the composition of those maps, which will take *s* to $x(u(s))$.

2. (Envelope) There is a submersion $\phi : \mathbb{R}^{n+1} \to \mathbb{R}$, $x \mapsto \phi(x)$, and the composition $\phi(x(s))$ of the map *x* with the map $\phi$ is constant for all *s* ; i.e., the point $x(s)$ always lies in the same level set of $\phi$.

The second definition will be more useful in what follows.

One defines the tangent vector field **t** (*s*) along the curve *x* (*s*) by:

$$\mathbf{t} = \frac{dx}{ds} = \frac{dx^i}{ds} \partial_i \tag{2.39}$$

When the curve parameter *s* is arc-length (measured from some reference point along the curve that defines $s = 0$), **t** will be a unit vector field. That follows from the definition of arc-length:

$$ds^2 \equiv \delta_{ij} \, dx^i \, dx^j = \left( \delta_{ij} \frac{dx^i}{ds} \frac{dx^j}{ds} \right) ds^2 = \| \mathbf{t} \|^2 \, ds^2. \tag{2.40}$$

We define the unit normal vector field $\mathbf{n}(s)$ to make to make the *curvature vector field* of the curve take the form:

$$\frac{d\mathbf{t}}{ds} = \frac{d^2 x}{ds^2} = \kappa \mathbf{n}, \qquad \kappa = \left\| \frac{d\mathbf{t}}{ds} \right\| = \left\| \frac{d^2 x}{ds^2} \right\|, \tag{2.41}$$

whenever the *curvature* $\kappa \neq 0$; otherwise $\mathbf{n}$ will be undefined; this will be the case when $x(s)$ describes a straight line or when the point $x(s)$ is an inflection point.

When $\kappa \neq 0$, $\mathbf{n}$ will be orthogonal to $\mathbf{t}$, since $\mathbf{t}$ is a unit vector:

$$0 = \frac{d}{ds}[\delta(\mathbf{t}, \mathbf{t})] = 2\delta\left(\mathbf{t}, \frac{d\mathbf{t}}{ds}\right) = 2\kappa\,\delta(\mathbf{t}, \mathbf{n}).$$

We have chosen the sign of $\kappa$ in order to be consistent with the usual Frenet-Serret equations for a spatial curve [**23**]. In a sense, curvature generalizes the centripetal acceleration of a curve of motion, as we shall see below.

Since $\mathbf{t}$ and $\mathbf{n}$ are not collinear (unless $\mathbf{n} = 0$; i.e., $\mathbf{t}$ is constant), they will span a plane in the tangent space at each point along the curve. That plane is called the *osculating plane* at that point because the circle of radius $1/\kappa$ (if it exists) with its center at the center of curvature "osculates" the curve at that point: i.e., they both share a common point and tangent vector at that point. In order to describe the changing of the osculating plane as $s$ varies (i.e., the "torsion" of the curve), one must then go to the third derivative of $x(s)$ with respect to $s$. If that third derivative vanishes (i.e., $\mathbf{n}$ is constant) then the curve will necessarily be planar. However, planar curves can still have non-vanishing third derivatives, such as circular or sinusoidal motions.

When one differentiates $\phi(x(s))$ with respect to $s$, one will get the condition on the tangent vector field to the curve:

$$N(\mathbf{t}) = d\phi(\mathbf{t}) = 0. \tag{2.42}$$

This can be expressed in component form by:

$$\frac{d\phi}{ds} = \frac{\partial \phi}{\partial x^i} \frac{dx^i}{ds} = 0. \tag{2.43}$$

In either form, one has that $\mathbf{t}$ must lie in the hypersurface that is annihilated by $N$ for every $s$. Of course, (2.43) also says that $\phi$ must be constant along the curve in question.

Since $N = d\phi$, we can also say that the rate of change of $N$ along the curve will be its Lie derivative with respect to $\mathbf{t}$:

$$\frac{dN}{ds} = L_{\mathbf{t}} N = i_{\mathbf{t}}\, d_\wedge N + d\, i_{\mathbf{t}} N = 0 + d\,(N(\mathbf{t})) = 0.$$

Hence, since $d_\wedge N$ vanishes, $N$ will be "constant" along a curve in a hypersurface. When we get to pseudo-hypersurfaces, we will see that it is no longer proper to assume that $d_\wedge N$ vanishes, and $N$ will not have to be constant along a curve in a pseudo-hypersurface, either.

A second differentiation of the constraint (2.42) on **t** with respect to $s$ will give a constraint on the curvature of the curve:

$$0 = \frac{dN}{ds}(\mathbf{t}) + N\left(\frac{d\mathbf{t}}{ds}\right) = dN(\mathbf{t},\mathbf{t}) + \kappa N(\mathbf{n}). \tag{2.44}$$

Since:

$$N(\mathbf{n}) = \delta(\mathbf{N}, \mathbf{n}) = \cos\theta, \tag{2.45}$$

where $\theta$ is the angle between **n** and **N** (or equivalently, the dihedral angle between the osculating plane of the curve and the normal section of the hypersurface that is spanned by **t** and **N**), equation (2.44) can be written in the form:

$$\kappa \cos\theta = -dN(\mathbf{t}, \mathbf{t}). \tag{2.46}$$

The left-hand side of this equation is the length of the projection of the curvature vector $\kappa$ **n** onto the normal section, which can be called the *normal curvature* $\kappa_n$, while the length of its projection onto the hyperplane $\Sigma_x$ – namely, $\kappa \sin\theta$ – can be called the *geodesic curvature* $\kappa_g$. Since **t** has no normal component, $dN(\mathbf{t}, \mathbf{t})$ will agree with $H(\mathbf{t},\mathbf{t})$, and one can then rewrite equation (2.46) in the form:

$$\kappa_n = -H(\mathbf{t}, \mathbf{t}). \tag{2.47}$$

*i. Distinguished curves in hypersurfaces.* – One can distinguish certain curves in hypersurfaces that basically relate to the eigenvectors of the second fundamental form.

For instance, one has *lines of curvature*, for which the tangent vector field **t** (*s*) represents an eigenvector of $H_b^a$ at each point $x$ (*s*). Note that the eigenvalue $\kappa_a$ (*s*) can itself vary along the curve, in general. When $\mathbf{v} \in \Sigma_x$ is an eigenvector of $H_b^a$, with eigenvalue $\kappa_v$, and $\mathbf{e}_a$ is an adapted frame for $\Sigma_x$, one will have:

$$H(\mathbf{v}, \mathbf{v}) = H_{ab} v^a v^b = g_{ac} H_b^c v^a v^b = \kappa_v g_{ac} v^a v^c = \kappa_v g(\mathbf{v}, \mathbf{v}). \tag{2.48}$$

When $H$ is hyperbolic (or restricted to a subspace of $\Sigma_x$ on which it is hyperbolic), an *asymptotic line* is a curve whose tangent vector is *isotropic*:

$$H(\mathbf{t}, \mathbf{t}) = 0 \tag{2.49}$$

at the point in question. (Such a vector cannot be a zero eigenvector, since a hyperbolic $H$ would not have zero eigenvalues.)

A curve $x(s)$ in a hypersurface $S$ is called a *normal curve* iff its normal vector field $\mathbf{n}(s)$ is collinear with the normal $\mathbf{N}(s)$ to $S$ for all $s$ ; that is then equivalent to the vanishing of the geodesic curvature at each point. Hence $\cos\theta = 1$, $\kappa_n = \kappa$, and one can then put the condition (2.47) into the form:

$$\kappa = -H(\mathbf{t}, \mathbf{t}) \,. \tag{2.50}$$

When one multiplies both sides of this by $\mathbf{n}$, one will obtain an expression for the constraint on the curvature of a normal curve in the form of:

$$\frac{d^2 x}{ds^2} = -H(\mathbf{t}, \mathbf{t})\,\mathbf{n} = -H(\mathbf{t}, \mathbf{t})\,\mathbf{N}, \tag{2.51}$$

which can be written in component form as:

$$\frac{d^2 x^i}{ds^2} = -C^i_{jk}\,\frac{dx^j}{ds}\frac{dx^k}{ds}, \tag{2.52}$$

in which, we have defined the third-rank tensor field:

$$C \equiv dN \otimes \mathbf{N}, \tag{2.53}$$

whose components are:

$$C^i_{jk} = \frac{\partial^2 \phi}{\partial x^j \partial x^k} N^i = \delta^{il}\,\frac{\partial^2 \phi}{\partial x^j \partial x^k}\frac{\partial \phi}{\partial x^l}\,. \tag{2.54}$$

Note that whereas $\mathbf{n}(s)$ is defined only along a curve, $\mathbf{N}$ is defined in all of $\mathbb{R}^{n+1}$. Hence, $C$ is defined everywhere, and is therefore independent of the choice of curve.

Although the form of (2.52) is strongly reminiscent of the usual geodesic equation, assuming that $C^i_{jk}$ are the components of a connection 1-form, this is not actually the case now. In particular, $C^i_{jk}$ are the components of a tensor field, while connection 1-forms are not actually tensorial. We can, however, think of the right-hand side of (2.52) as the *curvature due to the constraint*, since it is defined entirely by $d\phi$ and $d^2\phi$. In a sense, the curvature due to the constraint counteracts the curvature of the curve $x(s)$ itself.

*j. Path of shortest length.* – In the geometry of hypersurfaces, one can also consider a path $x(s)$ between two (sufficiently-close) points $x_0$ and $x_1$ on the hypersurface in question that is shorter than all of the possible ones, which is the original definition of a *geodesic*.

The problem of the shortest path between two points on a hypersurface is basically an elementary problem in the calculus of variations when we include constraints. Namely, we wish to find the extremum (viz., minimum, in the Euclidian case) of the path-length functional, which takes a path $x\,(s')$ between two points $x\,(0)$ and $x\,(s)$ in space to the length of that path:

$$s\,[x\,(s)] = \int_0^s ds = \int_0^s \sqrt{\delta_{ij}\frac{dx^i}{ds'}\frac{dx^i}{ds'}}\,ds', \tag{2.55}$$

subject to the constraint that the tangent vector must always lie in the hyperplane that is annihilated by the Pfaffian form $N$ at every point:

$$d\phi\left(\frac{dx}{ds'}\right) = \frac{d\phi}{ds'} = 0. \tag{2.56}$$

(The reason that we have switched to an arbitrary parameter $s'$ is that otherwise the integrand would be simply 1.)

When one introduces the Lagrange multiplier $\lambda\,(x)$, one can define the constrained Lagrangian:

$$\mathcal{L}(x,\dot{x}) = v(\dot{x}) + \lambda(x)\phi_{,i}\,\dot{x}^i, \tag{2.57}$$

in which the dot signifies the derivative with respect to $s'$ and we have set:

$$v(\dot{x}) = \frac{ds}{ds'} = \sqrt{\delta_{ij}\frac{dx^i}{ds'}\frac{dx^i}{ds'}}. \tag{2.58}$$

The arc-length functional $s\,[x\,(s)]$ will have an extremum iff its first variation:

$$\delta s\,[\delta x] = \int_0^s \left(\frac{\delta \mathcal{L}}{\delta x^i}\delta x^i\right)ds' \tag{2.59}$$

vanishes for all $\delta x$ when one assumes fixed endpoints and recalls that $(\delta_{ij}\,\dot{x}^i\,\dot{x}^j)^{1/2} = 1$. That will lead to the Euler-Lagrange equations for the extremal path $x\,(s)$:

$$0 = \frac{\delta \mathcal{L}}{\delta x^i} = \frac{d}{ds}\frac{\partial \mathcal{L}}{\partial \dot{x}^i} - \frac{\partial \mathcal{L}}{\partial x^i} = \frac{\delta v}{\delta x^i} + \frac{\delta}{\delta x^i}(\lambda\phi_{,i}\,\dot{x}^i) \tag{2.60}$$

Now:

$$\frac{\delta v}{\delta x^i} = \frac{d}{ds'}\frac{\partial v}{\partial \dot{x}^i} = \frac{d}{ds'}\left(\frac{1}{v}\delta_{ij}\dot{x}^j\right) = -\frac{\dot{v}}{v^2}\dot{x}_i + \frac{1}{v}\ddot{x}_i, \tag{2.61}$$

and when one reverts to $s' = s$, one will have $v = 1$, so the first term will vanish, and one will be left with:

$$\frac{\delta v}{\delta x^i} = \ddot{x}_i. \tag{2.62}$$

Meanwhile:

$$\frac{\delta}{\delta x^i}(\lambda \phi_{,j} \dot{x}^j) = \frac{d}{ds}(\lambda \phi_{,i}) - \frac{\partial(\lambda \phi_{,j})}{\partial x^i}\dot{x}^j = [\partial_j(\lambda \phi_{,i}) - \partial_i(\lambda \phi_{,j})]\dot{x}^j$$

$$= [(\partial_j \lambda)\phi_{,i} - (\partial_i \lambda)\phi_{,j}]\dot{x}^j + \lambda[\partial_j \phi_{,i} - \partial_i \phi_{,j}]\dot{x}^j$$

$$= \frac{d\lambda}{ds}\phi_{,i}.$$

That implies the following system of ordinary differential equations for $x(s)$:

$$\frac{d^2 x_i}{ds^2} = -\frac{d\lambda}{ds}\phi_{,i}. \tag{2.63}$$

We can also write (2.63) in intrinsic form:

$$\frac{dt}{ds} = -\frac{d\lambda}{ds} N, \tag{2.64}$$

in which the 1-form $t = t_i \, dx^i$ is the metric-dual to the tangent vector field **t**. When we compare this to (2.51), we see that we can set:

$$\frac{d\lambda}{ds} = H(\mathbf{t}, \mathbf{t}). \tag{2.65}$$

In any event, a curve for which (2.64) is true is clearly a normal curve, and we can summarize the result that we have obtained in a:

**Theorem:**

*A curve on a hypersurface is a normal curve iff it is a geodesic.*

**3. The geometry of pseudo-hypersurfaces.** – Many of the geometric constructions that were just made for hypersurfaces are based upon tangent and cotangent objects, which can exist independently of whether the hypersurface is represented as an embedded submanifold. Hence, one must consider its definition as an envelope, not a locus, and not assume that the envelope is completely integrable.

In particular, one can simply start with a Pfaff equation:

$$N = 0 \qquad (N = N_i \, dx^i), \tag{3.1}$$

and drop the assumption that the 1-form $N$ is exact (i.e., takes the form $d\phi$ for some smooth function $\phi$). Hence, one assumes that:

$$d_\wedge N = \tfrac{1}{2}(\partial_i N_j - \partial_j N_i)\, dx^i \wedge dx^j \neq 0. \tag{3.2}$$

*a. Definition of a pseudo-hypersurface.* – When the degree of integrability of the Pfaff equation $N = 0$ is greater than or equal to 1, but less than $n$, one calls the sub-bundle $\Sigma$ of the tangent bundle $T(\mathbb{R}^{n+1})$ that consists of all annihilating hyperplanes $\Sigma_x$ a *pseudo-hypersurface*. That is because it looks like the tangent bundle to a hypersurface, but the actual dimensions of the manifolds that are tangent to those hyperplanes are less than the dimension of the hyperplanes.

The differential of the normal 1-form $N$:

$$dN = \partial_i N_j \, dx^i \otimes dx^j \qquad (3.3)$$

is a second-rank, covariant tensor field on $\mathbb{R}^{n+1}$, but unlike $d^2\phi$ it is not generally symmetric. One can, however, polarize it into a symmetric and an anti-symmetric part:

$$dN = H + \omega, \qquad (3.4)$$

in which:

$$H = \tfrac{1}{2}(\partial_i N_j + \partial_j N_i) \, dx^i \, dx^j, \qquad \omega = d_\wedge N = \tfrac{1}{2}(\partial_i N_j - \partial_j N_i) \, dx^i \wedge dx^j, \qquad (3.5)$$

Since $dN$ has both symmetric and antisymmetric parts this time, it is no loss of generality to normalize $N$ to a unit covector field since any non-zero $N$ can be normalized, and any scalar multiple of $N$ will annihilate the same hyperplane at each point. From now on, we shall assume that **N** is a unit vector field.

Regardless of its degree of integrability, the 1-form $N$ can still be associated with a normal vector field **N** in the same way as in the integrable case; if $N$ is a unit covector field then **N** will be a unit vector field. If **N** were the flow velocity vector field for the motion of an extended material medium then $H$ would be the rate of strain tensor and $\omega$ would be the vorticity. However, as we shall see shortly, the symmetric part $H$ of $dN$ (or more precisely, its restriction to the annihilating hyperplanes $\Sigma_x$) has many of the same uses as the second fundamental form for a hypersurface.

*b. Inclusion and projection operators.* – Since we no longer have the luxury of an embedding of a submanifold whose tangent spaces will agree with the hyperplanes in $\Sigma$, we will have to make do with the hyperplanes themselves as local exemplars of tangent spaces and project the general tangent vector $\mathbf{v} \in T_x$ onto the hyperplane $\Sigma_x$.

To begin with, in the absence of an embedding $x : S \to \mathbb{R}^{n+1}$ that would define a linear injection $dx|_x : T_u S \to T_x$ whose image is the hyperplane $\Sigma_x$, we find that we can use the linear injection $\iota_x : \Sigma_x \to T_x$, $\mathbf{v} \mapsto \mathbf{v}$, which is referred to as the *inclusion map* for the subspace $\Sigma_x$; it is basically a restriction of the identity map to the subspace.

Since we are assuming that our ambient space $\mathbb{R}^{n+1}$ has a metric on it, the covector field (i.e., 1-form) $N$ can be associated with a vector field **N**, which then generates a line $[\mathbf{N}]_x$ in each tangent space $T_x$ that is orthogonal to the hyperplane $\Sigma_x$. That also leads to a direct-sum decomposition:

$$T_x = \Sigma_x \oplus [\mathbf{N}]_x,$$

so every tangent vector $\mathbf{v} \in T_x$ will admit a unique decomposition

$$\mathbf{v} = \mathbf{v}_t + \mathbf{v}_n, \tag{3.6}$$

where $\mathbf{v}_t$ is the tangential component, which lies in $\Sigma_x$, and $\mathbf{v}_n$ is the normal component, which lies along $[\mathbf{N}]_x$. In particular, we can use $N$ and $\mathbf{N}$ to define both components:

$$\mathbf{v}_n = N(\mathbf{v})\mathbf{N}, \quad \mathbf{v}_t = \mathbf{v} - \mathbf{v}_n. \tag{3.7}$$

This also allows us to define projection operators $P_t : T_x \to \Sigma_x$, $\mathbf{v} \mapsto \mathbf{v}_t$ and $P_n : T_x \to [\mathbf{N}]_x$, $\mathbf{v} \mapsto \mathbf{v}_n$, which can then be expressed in the form:

$$P_n = N \otimes \mathbf{N}, \quad P_t = I - P_n = I - N \otimes \mathbf{N}, \tag{3.8}$$

in which $I$ is the identity operator on $T_x$.

The component forms of these operators in terms of the natural frame and coframe on $\mathbb{R}^{n+1}$ are:

$$[P_n]^i_j = N_j N^i, \qquad [P_t]^i_j = \delta^i_j - N_j N^i. \tag{3.9}$$

Furthermore, when one composes the inclusion of $\Sigma_x$ in $T_x$ with the projection of $T_x$ onto $\Sigma_x$, one will get the identity on $\Sigma_x$, while the opposite composition will still be the projection onto $\Sigma_x$. The composition of the inclusion with the normal projection will vanish in either order:

$$P_t \cdot \iota = I_\Sigma, \qquad \iota \cdot P_t = P_t, \qquad P_n \cdot \iota = \iota \cdot P_n = 0. \tag{3.10}$$

One also has a dual decomposition of the cotangent space $T_x^*$ into:

$$T_x^* = \Sigma_x^* \oplus [N]_x$$

and a corresponding unique decomposition of any covector $\alpha$ in $T_x^*$ into:

$$\alpha = \alpha_t + \alpha_n, \quad \alpha_t \in \Sigma_x^*, \qquad \alpha_n \in [N]_x. \tag{3.11}$$

However, since the transpose of the projection $P_t : T_x \to \Sigma_x$ is a linear map $P_t^* : \Sigma_x^* \to T_x^*$, which is then the inclusion of $\Sigma_x^*$ as a linear subspace of $T_x^*$, we see that if we want to go the opposite direction and project elements of $T_x^*$ onto $\Sigma_x^*$ then we will need a different map from $P_t^*$. We can then use the transpose of the inclusion map $\iota_x^* : T_x^* \to \Sigma_x^*$ to convert covectors on $\mathbb{R}^{n+1}$ into covectors on $\Sigma_x$.

One can then project an arbitrary-rank tensor at $x$:

$$T = T^{i_1 \cdots i_r}_{j_1 \cdots j_s} \partial_{i_1} \otimes \cdots \otimes \partial_{i_r} \otimes dx^{j_1} \otimes \cdots \otimes dx^{j_s}$$

onto a corresponding tensor in $\Sigma_x$ by applying $P_t$ to the vectors in the tensor product and $\iota_x^*$ to the covectors:

$$\overline{T} = T^{i_1 \cdots i_r}_{j_1 \cdots j_s} P_t(\partial_{i_1}) \otimes \cdots \otimes P_t(\partial_{i_r}) \otimes \iota_x^*(dx^{j_1}) \otimes \cdots \otimes \iota_x^*(dx^{j_s}).$$

This will allow one to re-express $\overline{T}$ in terms of the natural frame and coframe on $\mathbb{R}^{n+1}$ with new projected components:

$$\overline{T} = \overline{T}^{i_1 \cdots i_r}_{j_1 \cdots j_s} \partial_{i_1} \otimes \cdots \otimes \partial_{i_r} \otimes dx^{j_1} \otimes \cdots \otimes dx^{j_s}, \qquad (3.12)$$

in which:

$$\overline{T}^{i_1 \cdots i_r}_{j_1 \cdots j_s} = T^{k_1 \cdots k_r}_{l_1 \cdots l_s} P^{i_1}_{k_1} \cdots P^{i_r}_{k_r} \iota^{l_1}_{j_1} \cdots \iota^{l_s}_{j_s}.$$

Since $P_t(\mathbf{v}) = \mathbf{v}$ and $\iota(\mathbf{v}) = \mathbf{v}$ for any tangent vector $\mathbf{v} \in \Sigma$, the projected tensor $\overline{T}$ is really just the restriction of the tensor $T$ to vectors in $\Sigma$ and covectors in $\Sigma^*$.

*c. Adapted frames.* – The concept of a tangent frame at $T_x$ being adapted to the direct sum decomposition $\Sigma_x \oplus [\mathbf{N}]_x$ does not change from the completely-integrable case, nor does that of a coframe that is adapted to the dual decomposition of $T_x^*$. What does change is that without complete integrability, one cannot speak of adapted coordinate systems anymore.

For an adapted frame $\mathbf{e}^{n+1} = \mathbf{N}$, $\theta^{n+1} = N$, so the matrix of the inclusion map $\iota_x$ in an adapted frame will be:

$$[\iota]^i_j = \left[\begin{array}{c|c} \delta^a_b & 0 \\ \hline 0 & 0 \end{array}\right], \qquad (3.13)$$

which can be abbreviated to:

$$[\iota]^i_b = \left[\begin{array}{c} \delta^a_b \\ \hline 0 \end{array}\right], \qquad (3.14)$$

since the normal component of any vector in $\Sigma_x$ will be zero in an adapted frame.

The component matrices for $P_n$ and $P_t$ will take the forms:

$$[P_n]^i_j = \left[\begin{array}{c|c} 0 & 0 \\ \hline 0 & 1 \end{array}\right], \qquad [P_t]^i_j = \left[\begin{array}{c|c} \delta^a_b & 0 \\ \hline 0 & 0 \end{array}\right], \qquad (3.15)$$

respectively, so the matrix of the projection operator $P_t$ can be abbreviated to:

$$[P_t]^a_j = \delta^a_j = \left[\begin{array}{c|c} \delta^a_b & 0 \end{array}\right], \qquad (3.16)$$

which is the transpose of $[\iota]^i_b$.

Note that the compositions in (3.10) become matrix compositions, which say simply that:

$$\delta_j^a \delta_b^j = \delta_b^a, \quad \delta_a^i \delta_j^a = [P_t]_j^i, \quad [P_n]_k^i [\iota]_j^k = [\iota]_k^i [P_n]_j^k = 0 \tag{3.17}$$

in an adapted frame.

For an adapted frame, the inclusion and projection matrices will simply preserve the tangent members of the frame and annihilate the normal one, and the sum of terms on the right-hand side of (3.12) will collapse to only the terms that do not include a factor of $\mathbf{e}_{n+1}$ or $\theta^{n+1}$ in the tensor product:

$$T = \overline{T}_{b_1 \cdots b_s}^{a_1 \cdots a_r} \mathbf{e}_{a_1} \otimes \cdots \otimes \mathbf{e}_{a_r} \otimes \theta^{b_1} \otimes \cdots \otimes \theta^{b_s}, \tag{3.18}$$

in which:

$$\overline{T}_{b_1 \cdots b_s}^{a_1 \cdots a_r} = T_{j_1 \cdots j_s}^{i_1 \cdots i_r} \delta_{i_1}^{a_1} \cdots \delta_{i_r}^{a_r} \delta_{b_1}^{j_1} \cdots \delta_{b_s}^{j_s}.$$

Hence, the components $\overline{T}_{b_1 \cdots b_s}^{a_1 \cdots a_r}$ are now simply a subset of the components $T_{j_1 \cdots j_s}^{i_1 \cdots i_r}$.

Although coordinate systems that are adapted to both summands in the decomposition $\Sigma_x \oplus [\mathbf{N}]_x$ do not exist when $\Sigma$ is not completely integrable, one can, however, have a coordinate system $(x^1, \ldots, x^{n+1})$ that is adapted to just the normal vector field $\mathbf{N}$; i.e., the natural vector field $\partial_{n+1}$ is collinear with $\mathbf{N}$. One must simply keep in mind that tangent spaces to the level hypersurfaces of the $(n+1)^{\text{th}}$ coordinate $x^{n+1}$ cannot coincide with $\Sigma_x$.

*d. The first fundamental form of a pseudo-hypersurface.* – An example of the process of projecting tensor fields is how one might project the metric tensor field $\delta$ on $\mathbb{R}^{n+1}$ onto a corresponding metric $g$ on $\Sigma$ by using the projection $\iota_x^*: T_x^* \to \Sigma_x^*$ that was defined above.

Hence, we get a projected metric tensor field on $\Sigma$ by way of:

$$g = g_{ij} dx^i dx^j, \tag{3.19}$$

with:

$$g_{ij} = \delta_{kl} [\iota_x^*]_i^k [\iota_x^*]_j^l. \tag{3.20}$$

In an adapted coframe $\theta^a$, (3.19) will become:

$$g = \delta_{ab} \theta^a \theta^b, \tag{3.21}$$

which is simply the restriction of $\delta$ to the hyperplanes of $\Sigma$.

*e. Second fundamental form.* – For a hypersurface, the second fundamental form was defined to be the pull-back $x^* dN = x^* d^2\phi$ of the differential of the normal covector field along the embedding of the hypersurface. When one drops the assumption of complete integrability, no such embedding will exist any more, but one can substitute the linear inclusion $\iota: \Sigma \to T(\mathbb{R}^{n+1})$ for the differential map to the embedding and define the pull-back of the differential of the normal 1-form to a second-rank covariant tensor field $\iota^* dN$ on $\Sigma$. (Really, all that one is doing in this case is restricting the tensor field $dN$ on $\mathbb{R}^{n+1}$ to the vectors in the hyperplanes of $\Sigma$.)

We can also pull back the decomposition (3.4) to a decomposition of $\iota^* dN$:

$$\iota^* dN = \bar{H} + \iota^* \omega, \qquad (3.22)$$

whose symmetric part is:

$$\bar{H} = \iota^* H = H_{ab}\, \theta^a\, \theta^b. \qquad (3.23)$$

Hence, we are simply truncating the components of $H_{ij}$ in that adapted coframe.

The symmetric, second-rank, covariant tensor field $\bar{H}$ on $\Sigma$ is what we shall call the *second fundamental form* of the pseudo-hypersurface that is defined by $N = 0$, although some (e.g., Wundheiler [**21**]) reserve that term for $\iota^* dN$. Our justification for that restriction is that when we get to distinguished curves, such as normal curves, the only part of $\iota^* dN$ that will bear upon the nature of such things is the symmetric part, and it is easier to talk about eigenvalues of symmetric matrices than it is in the general case.

*f. Eigenvalues of $\bar{H}$.* – Since $H_{ab}$ is symmetric in $a$ and $b$, the matrix:

$$H^a_{\ b} = g^{ac}\, H_{cb} \qquad (3.24)$$

will have real eigenvalues $\kappa_a$, $a = 1, \ldots, n$ (which do not have to be non-zero) and orthogonal eigenvectors.

Hence, in a principal frame for $\bar{H}$ (which is composed of eigenvectors), one will have:

$$H^a_{\ b} = \text{diag}\,[\kappa_1, \ldots, \kappa_n]. \qquad (3.25)$$

Since a principal frame can be rescaled to be orthonormal, and the first fundamental form $g$ has the components $\delta_{ab}$ with respect to *any* orthonormal frame, one sees that a principal frame for $H$ will also diagonalize $g_{ab}$.

The eigenvalues $\kappa_a$, $a = 1, \ldots, n$ of $\bar{H}$ at each point in $\mathbb{R}^{n+1}$ are called the *principal curvatures* of the pseudo-hypersurface $\Sigma$ and play essentially the same role as the principal curvatures of a hypersurface. In particular, when one holds $\kappa$ constant, the equation:

$$\kappa = \bar{H}(\mathbf{t}, \mathbf{t}) = H_{ab}\, t^a\, t^b, \qquad a, b = 1, \ldots, n \qquad (3.26)$$

will define a quadric hypersurface in each hyperplane $\Sigma_x$, which will be called a *fundamental quadric for $\bar{H}$* at $x$; when $\kappa$ is a principal curvature, it will be called a *principal quadric*.

One can then classify the points in $\mathbb{R}^{n+1}$ according to the signature type of $\bar{H}$ as a quadratic form on tangent vectors with the same terminology as in the case of hypersurfaces. Hence, points can once more be elliptic, hyperbolic, parabolic, cylindrical, and umbilic as before, and one can decompose the hyperplanes $\Sigma_x$ into direct sums of eigenspaces this time. In the general case, one will have $\Sigma_x = \Sigma_x^- \oplus \Sigma_x^0 \oplus \Sigma_x^+$, which correspond to vectors with negative, zero, and positive eigenvalues, respectively.

*g. Invariants of* $\bar{H}$. – One has the same two frame-invariants for $\bar{H}$ as before, namely, the trace of $\bar{H}$:

$$\operatorname{Tr} \bar{H} = H_a^a = \sum_{a=1}^{n} \kappa_a \qquad (3.27)$$

which defines the *mean curvature* of the pseudo-hypersurface:

$$\bar{\kappa} = \frac{1}{n} \operatorname{Tr} \bar{H}, \qquad (3.28)$$

and the determinant:

$$K = \det \bar{H} = \prod_{a=1}^{n} \kappa_a, \qquad (3.29)$$

which defines its *Gaussian curvature*.

*h. Curves in pseudo-hypersurfaces.* – A smooth curve $x : \mathbb{R} \to \mathbb{R}^{n+1}$, $s \mapsto x(s)$, where $s$ is the arc-length parameter, is said to *lie in the pseudo-hypersurface* $\Sigma$ that is defined by a Pfaff equation $N = 0$ iff its tangent vector $\mathbf{t}(s)$ belongs to $\Sigma_x(s)$ for all $s$; hence:

$$N(\mathbf{t}(s)) = N_i \frac{dx^i}{ds} = 0 \qquad \text{for all } s. \qquad (3.30)$$

This time, since when $N$ does not generally take the form $d\phi$, we can no longer characterize this condition by saying that $\phi$ is constant along the curve.

Similarly, when we look at how $N$ itself varies along the curve, we will get:

$$\mathbf{L_t} N = \mathbf{L_t} d_\wedge N = i_\mathbf{t} \omega, \qquad (3.31)$$

which no longer has to vanish, since $\omega$ will decompose into:

$$\omega = \varpi + \eta \wedge N, \qquad (3.32)$$

in which the 2-form $\varpi$ and the 1-form $\eta$ are defined over $\Sigma$, so:

$$i_\mathbf{t} \omega = i_\mathbf{t} \varpi + \eta(\mathbf{t}) N. \qquad (3.33)$$

By differentiating the constraint (3.30) on **t** with respect to *s*, one can get a corresponding constraint on the curvature of the curve *x* (*s*):

$$\frac{dN}{ds}(\mathbf{t}) + N\left(\frac{d\mathbf{t}}{ds}\right) = dN(\mathbf{t},\mathbf{t}) + N\left(\frac{d\mathbf{t}}{ds}\right) = \bar{H}(\mathbf{t},\mathbf{t}) + \kappa N(\mathbf{n}) = 0,$$

since the evaluation of the antisymmetric part of *dN* on the pair of tangent vectors (**t**, **t**) – namely, $\omega(\mathbf{t}, \mathbf{t})$ – vanishes due to that antisymmetry and $H(\mathbf{t}, \mathbf{t}) = \bar{H}(\mathbf{t},\mathbf{t})$.

The condition on the curvature above (2.46) now takes the form:

$$\kappa_n = -\bar{H}(\mathbf{t}, \mathbf{t}), \tag{3.34}$$

in which the normal curvature $\kappa_n$ is defined as before. Hence, we have simply replaced the $d^2\phi$ (which exists only in the integrable case) with *H* and restricted it to $\Sigma$ to get $\bar{H}$.

*i. Distinguished curves in pseudo-hypersurfaces.* – One can distinguish certain curves in pseudo-hypersurfaces in the same way that one does for hypersurfaces. The *lines of curvature* are the ones for which the tangent vector field **t** (*s*) represents an eigenvector of the matrix:

$$\bar{H}^a_b = g^{ac}\bar{H}_{cb} \tag{3.35}$$

at each point *x* (*s*). Note that the eigenvalue $\kappa_a$ (*s*) can itself vary along the curve, in general.

When $\bar{H}$ is hyperbolic, an *asymptotic line* is once more a curve whose tangent vector is *isotropic*:

$$\bar{H}(\mathbf{t}, \mathbf{t}) = 0 \tag{3.36}$$

at the point in question.

A curve that lies in the pseudo-hypersurface $\Sigma$ is once more called a *normal curve* iff **n** is collinear with **N**; i.e. **n** = **N**, That will make *N* (**n**) = 1, so the condition (3.34) will become:

$$\kappa = -\bar{H}(\mathbf{t}, \mathbf{t}) = -H_{ij}\frac{dx^i}{ds}\frac{dx^j}{ds}. \tag{3.37}$$

If we recall the definition (2.41) of the curvature vector then multiplying both sides of the first of these two equations by **n** will give:

$$\frac{d^2x}{ds^2} = -\bar{H}(\mathbf{t}, \mathbf{t})\,\mathbf{n} \quad \text{or} \quad \frac{d^2x^i}{ds^2} = -C^i_{jk}\frac{dx^j}{ds}\frac{dx^k}{ds}, \tag{3.38}$$

in which we have defined the third-rank tensor $C \equiv H \otimes N$ whose components are:

$$C^i_{jk} = N^i H_{jk} = \tfrac{1}{2}N^i(\partial_j N_k + \partial_k N_j), \tag{3.39}$$

which might be compared with the previous definition (2.54) in the integrable case.

*j. Shortest path in a pseudo-hypersurface.* – The question of geodesics of Pfaffians has addressed by numerous researchers along the way, such as von Lilienthal [**13**], Synge [**16**], Schouten [**18**], and Franklin and Moore [**20**]. Although one might well question whether the concept of an action functional will still be applicable when one does not have holonomic constraints (see Lanczos [**29**] or Sommerfeld [**30**]), we shall proceed naively in the manner of those distinguished figures before us.

Basically, one starts by forming the same arc-length functional *s* [*x*] as before in the completely-integrable case of a hypersurface − namely, (2.55) – and introducing the constraint by way of a Lagrange multiplier. That will once more produce the Lagrangian (2.57), and one will again have:

$$\frac{\delta v}{\delta x^i} = \ddot{x}_i , \qquad (3.40)$$

but this time, when one replaces $\phi_{,i}$ with $N_i$, one will now have:

$$\begin{aligned}\frac{\delta}{\delta x^i}(\lambda N_j \dot{x}^j) &= \frac{d}{dt}(\lambda N_i) - \frac{\partial(\lambda N_j)}{\partial x^i}\dot{x}^j = [\partial_j(\lambda N_i) - \partial_i(\lambda N_j)]\dot{x}^j \\ &= [(\partial_j \lambda) N_i - (\partial_i \lambda) N_j]\dot{x}^j + \lambda[\partial_j N_i - \partial_i N_j]\dot{x}^j \\ &= \frac{d\lambda}{ds}N_i + \lambda[\partial_j N_i - \partial_i N_j]\dot{x}^j,\end{aligned}$$

Hence, the Euler-Lagrange equations will now take the form:

$$\frac{dt}{ds} = -\frac{d\lambda}{ds}N - \lambda i_{\mathbf{t}} d_\wedge N , \qquad (3.41)$$

which has picked up a contribution from the non-vanishing of $d_\wedge N$, which would have vanished before in (2.64). When we take the exterior product of both sides with *N*, we will get:

$$N \wedge \frac{dt}{ds} = -\lambda N \wedge i_{\mathbf{t}} d_\wedge N = \lambda i_{\mathbf{t}}(N \wedge d_\wedge N), \qquad (3.42)$$

and we will see that when we drop the assumption of complete integrability, we can no longer infer that *dt* / *ds* must be collinear with *N*. Hence, we can no longer say that normal curves must coincide with geodesics for a pseudo-hypersurface, since the curvature of the geodesic in the present case includes not only a normal component, but possibly a tangent component that arises from the non-integrability of the Pfaff equation *N* = 0. However, we can say that a normal curve on a pseudo-hypersurface is a geodesic for which $i_{\mathbf{t}} d_\wedge N$ is a normal 1-form.

Since $d_\wedge N$ will take the general form:

$$d_\wedge N = \eta \wedge N + \varpi,$$

in which $\eta$ and $\varpi$ are a purely tangential 1-form and 2-form, respectively, that will make:

$$i_{\mathbf{t}}\, d_\wedge N = \eta\,(\mathbf{t})\, N + i_{\mathbf{t}}\, \varpi.$$

Since $i_{\mathbf{t}}\, \varpi$ is a tangential 1-form, the only way that $i_{\mathbf{t}}\, d_\wedge N$ can be normal is if $i_{\mathbf{t}}\varpi$ vanishes. If $\{\theta^a,\, a = 1, \ldots, n\}$ is adapted to $\Sigma$ then one can write $\varpi = \tfrac{1}{2} \varpi_{ab}\, \theta^a \wedge \theta^b$, which will make:

$$i_{\mathbf{t}}\varpi = (\varpi_{ab}\, t^a)\, \theta^b,$$

which will vanish iff $\varpi_{ab}\, t^a$ does, or equivalently $\varpi_{ab}\, t^b$.

Now, the matrix $\varpi^a_b = g^{ac}\varpi_{ab}$ represents an infinitesimal rotation in the hyperplane $\Sigma_x$. If $\mathbf{t}$ lies along its rotational axis then it is possible for $\varpi^a_b\, t^b$ to vanish when $\varpi^a_b$ does not as long as the dimension of $\Sigma_x$ is greater than two. We will then find the:

**Theorem:**

*A geodesic in a pseudo-hypersurface whose tangent vector field is $\mathbf{t}\,(s)$ will be a normal curve iff $\mathbf{t}$ is the axis of rotation for the infinitesimal rotation $\varpi^a_b$.*

That implies the:

**Corollary:**

*A geodesic on a pseudo-surface will be a normal curve iff the pseudo-surface is a surface.*

That is because when $n = 2$ (i.e., $\Sigma$ is a pseudo-surface), the vanishing of $i_{\mathbf{t}}\varpi$ will imply the vanishing of $\varpi$, which will make:

$$d_\wedge N = \eta \wedge N,$$

which will imply that:

$$N \wedge d_\wedge N = 0\,;$$

i.e., the pseudo-surface must be a surface.

Therefore, in general, the condition that $i_{\mathbf{t}}\, \varpi$ must vanish typically only singles out a class of curves whose tangent vector fields have the specified property in relation to $d_\wedge N$.

**4. Kinematics in pseudo-hypersurfaces.** – As long as one is dealing with the kinematics of point-like matter that moves in $\mathbb{R}^{n+1}$, one can define the motion of such a point by a smooth curve in $\mathbb{R}^{n+1}$. If one also wishes that the motion should be constrained by a condition that defines a pseudo-hypersurface $\Sigma$ (namely, that its tangent vectors should always lie in certain hyperplanes $\Sigma_x$) then one can restrict oneself to smooth curves that lie in the pseudo-hypersurface in the sense that was defined above.

*a. Kinematics of curves.* – The main alteration to the considerations that were made above is that the curve parameter must now be time *t*, rather than arc-length *s*. As long as the change of parameter is an orientation-preserving diffeomorphism of $\mathbb{R}$ that takes *s* to *s*(*t*), one can define the *speed* of the parameterization *s* to be:

$$v(t) = \left.\frac{ds}{dt}\right|_t . \tag{4.1}$$

Because one assumes that the reparameterization is an orientation-preserving diffeomorphism, this real number must be greater than zero for all *t* ; i.e.:

$$v(t) > 0 \qquad \text{for all } t. \tag{4.2}$$

Hence, the *velocity* **v** (*s*) of the curve *x* (*t*) can be expressed in terms of its unit-tangent vector field **t** (*s*) by way of:

$$\mathbf{v}(t) = \left.\frac{dx}{dt}\right|_t = \left.\frac{ds}{dt}\right|_t \left.\frac{dx}{ds}\right|_{s(t)} = v(t)\,\mathbf{t}(s(t)). \tag{4.3}$$

Although this represents a simple rescaling of the unit tangent vector at each point along the curve that preserves the direction of the tangent, it does not imply a mere rescaling of the acceleration. Indeed, another differentiation by *t* will yield:

$$\mathbf{a}(t) = \left.\frac{d\mathbf{v}}{dt}\right|_t = \left.\frac{dv}{dt}\right|_t \mathbf{t}(s(t)) + v(t)\left.\frac{d\mathbf{t}}{ds}\right|_{s(t)} = \dot{v}\mathbf{t} + \kappa v^2 \mathbf{n}. \tag{4.4}$$

The last expression clearly describes the well-known decomposition of an acceleration vector into a tangential component $a_t$, whose magnitude is:

$$a_t = \dot{v} \tag{4.5}$$

and a "radial" component that points towards the instantaneous center of rotation, and whose magnitude is:

$$a_c = -\kappa v^2 = -\frac{v^2}{r}, \tag{4.6}$$

when one introduces the *instantaneous radius of curvature* $r = 1 / \kappa$. The last expression for $a_c$ is clearly the instantaneous centripetal acceleration of the motion, but one sees that the first expression (viz., $-\kappa v^2$) also shows that centripetal acceleration is essentially a rescaling of the curvature of the curve of motion.

One sees that even though the acceleration vector is not typically collinear with the normal vector, nonetheless, unless the acceleration is collinear with the velocity, the velocity vector and the acceleration vector still span the same osculating plane as **t** and **n**. Once again, in order to get a non-planar motion, the curve $x$ ($t$) must have a non-vanishing third derivative (viz., a "jerk").

In the case of *uniform motion*, for which $\dot{v} = 0$ (i.e., constant speed), the acceleration itself becomes a rescaling of the normal vector field to the curve. The general form of the transformation from $t$ to $s$ will then be an affine transformation:

$$s\ (t) = s\ (0) + v\ t. \tag{4.7}$$

*b. Types of constrained motion.* – Typically, motion is constrained by requiring the velocity vector to lie in some specified submanifold of the tangent space at each point along the curve. The submanifold can be linear, affine, or nonlinear, and the dimension of the submanifold defines the number of *degrees of freedom* of the constrained motion.

An ($n - p + 1$)-dimensional linear subspace of an ($n + 1$)-dimensional tangent space is typically defined by the vanishing a finite set of non-zero 1-forms $N^a$, $a = 1, \ldots, p$. When those 1-forms are defined everywhere on $\mathbb{R}^{n+1}$, the resulting exterior differential system:

$$N^a = 0 \tag{4.8}$$

becomes a *Pfaffian system*. The question of its degree of integrability then becomes quite involved, and typically centers around the "Cartan-Kähler theorem," which is, in turn, based in the Cauchy-Kovalevskaya theorem for systems of partial differential equations. We shall not deal with the question of the integrability of Pfaffian systems here, but focus on only the case where $p = 1$, which defines the Pfaff equation ([1]). One also notes that it is typically no loss of generality to assume that the 1-forms $N^a$ are linearly independent, or even orthonormal.

The main distinction that we need to make here is between *holonomic* constraints, for which the Pfaffian system is completely integrable, and *non-holonomic* constraints, for which it is not ([2]).

In component form, the constraint on the velocity vector **v** that is defined by (4.8) will be:
$$N^a_i\ v^i = 0. \tag{4.9}$$

When $p = 1$, that will become the constraint that we have been working with all along.

An affine submanifold is defined when the right-hand side of (4.9) is non-zero:

$$N^a_i\ v^i = b^a. \tag{4.10}$$

---

([1]) Readers who wish to go further in that direction might look at the original works by Cartan [**31**] and Kähler [**32**], as well as the more modern treatment in Bryant, Chern, et al [**33**].

([2]) One should be cautioned that in general relativity, it is more customary to negate the word "holonomic" with *anholonomic*.

The most common nonlinear constraints are defined by systems of algebraic equations:

$$\mathsf{P}^a [v^i] = b^a, \tag{4.11}$$

such as when $\mathsf{P}^a [v^i]$ is a polynomial in the $v^i$. For instance, light rays are basically geodesics in space that are constrained by the quadratic constraint that $g(\mathbf{v}, \mathbf{v}) = 0$, where $g$ is the Lorentzian metric on the space-time manifold.

One can further classify constraints as being time-dependent – or *rheonomic* – or time-independent – or *scleronomic*. When the ambient space includes time, such as in relativistic mechanics, the distinction is not as essential, so we shall typically restrict ourselves to the time-independent case.

*c. Kinematics of constrained motion.* – If one differentiates the linear constraint (4.9) on $\mathbf{v}$ ($p = 1$), namely:

$$N(\mathbf{v}) = 0, \tag{4.12}$$

with respect to $t$ then one will get:

$$0 = \frac{dN}{dt}(\mathbf{v}) + N(\mathbf{a}) = dN(\mathbf{v}, \mathbf{v}) + \dot{v} N(\mathbf{t}) + \kappa v^2 N(\mathbf{n}) = H(\mathbf{v}, \mathbf{v}) + \kappa v^2 N(\mathbf{n})$$

or

$$\kappa v^2 N(\mathbf{n}) = - H(\mathbf{v}, \mathbf{v}). \tag{4.13}$$

If one replaces $\kappa N(\mathbf{n})$ with $\kappa_n$ and replaces $H$ with $\bar{H}$ (since $\mathbf{v} \in \Sigma$) then this will take the form:

$$\kappa_n v^2 = - \bar{H}(\mathbf{v}, \mathbf{v}). \tag{4.14}$$

When one replaces $\mathbf{v}$ with $v \mathbf{t}$, this will be simply a rescaling of the constraint (2.46) on the normal vector field that was defined above by the scaling factor $v^2$.

Since a rescaling of $\bar{H}$ will not change its eigenvectors, but only its eigenvalues, the distinguished curves that $\bar{H}$ defines will still trace out the same points as the lines of curvature and asymptotic lines that one gets from using an arc-length parameterization. Umbilic points will still be umbilic points, and Dupin indicatrix will be defined the same way, but the vectors that solve it will simply be rescaled by $1 / v$ from the velocity vectors.

Recall that a curve in the pseudo-hypersurface $\Sigma$ was previously defined to be a normal curve iff its curvature was proportional to $\mathbf{N}$. From (4.4), that will now take the form of saying that the motion must be uniform ($\dot{v} = 0$) and the centripetal acceleration $\mathbf{a}_c = - \kappa v^2 \mathbf{n}$ must be collinear with $\mathbf{N}$. The condition (4.13) can then be written:

$$a_c = \bar{H}(\mathbf{v}, \mathbf{v}), \tag{4.15}$$

Since $\mathbf{a}$ coincides with $-\mathbf{a}_c$, multiplying both sides of (4.15) by $\mathbf{n} = \mathbf{N}$ will give:

$$\mathbf{a} = -\bar{H}(\mathbf{v},\mathbf{v})\mathbf{N}, \qquad (4.16)$$

which can be expressed in component form as:

$$\frac{d^2 x^i}{dt^2} = -C^i_{jk}\frac{dx^j}{dt}\frac{dx^k}{dt}, \qquad (4.17)$$

in which $C^i_{jk}$ is defined in the same way as before. Thus, the tensor $C$ is once more defined on all of $\mathbb{R}^{n+1}$, and not just a chosen curve.

Hence, the main difference between defining a normal curve in a pseudo-hypersurface that is parameterized by arc-length and defining one that is parameterized by time is that in the latter case, one must include the further constraint that the motion must be uniform in order to arrive at an equation of the form (4.17).

When one repeats the calculations that lead up to the equation of a geodesic that was constrained to lie in a pseudo-hypersurface with the curve parameter $s' = t$ this time, one will get:

$$\frac{\delta v}{\delta \dot{x}^i} = \frac{d}{dt}\left(\frac{1}{v}\dot{x}_i\right) = \frac{1}{v}(\ddot{x}_i - \dot{v}t_i) = \frac{1}{v}(\kappa v^2 n_i) = \kappa v\, n_i = v\frac{d^2 x_i}{ds^2}$$

and

$$\frac{\delta}{\delta \dot{x}^i}(\lambda N_j \dot{x}^j) = [\partial_j(\lambda N_i) - \partial_i(\lambda N_j)]\dot{x}^j = \frac{d\lambda}{dt}N + \lambda i_{\dot{x}}\omega.$$

Hence, the equation for geodesic trajectories on a pseudo-hypersurface becomes:

$$\kappa n_i = -\frac{d\lambda}{ds}N_i - \lambda(i_t\omega)_i, \qquad (4.18)$$

which is (3.41) again, and that can also be written as:

$$\boxed{\ddot{x} = \dot{v}t - v\frac{d\lambda}{dt}N - v\lambda(i_{\dot{x}}\omega).} \qquad (4.19)$$

Since $\dot{v}$ vanishes for a normal curve, we have the:

**Theorem:**

*In order for a geodesic on a pseudo-hypersurface to be a normal curve, it is necessary and sufficient that it must have constant speed and $i_{\dot{x}}\omega$ must be normal or vanish.*

This gives the revised:

**Theorem:**

*A geodesic on a pseudo-hypersurface will be a normal curve iff it has constant speed and* **t** *is the axis of rotation of the infinitesimal rotation* $\omega_b^a$.

As before, when $n = 2$, we will have the:

**Corollary:**

*A geodesic on a pseudo-surface will be a normal curve iff it has constant speed and the pseudo-surface is a surface.*

**5. Example: charge moving in an electromagnetic field.** – Although the theory of electromagnetism is typically treated as a gauge field theory (see, e.g., [**34**]) in which the gauge group is $U(1)$, for the present purposes, only a few aspects of that picture will be necessary. Thus, a deep understanding of the theory of connections on principal fiber bundles will not be as necessary as one might assume.

In general, our usual background space of $\mathbb{R}^{n+1}$ will be a five-dimensional differentiable manifold $P$ that would typically be regarded as the total space of a $U(1)$-principal bundle $\pi: P \to M$ over the four-dimensional space-time manifold $M$, which will be given a Lorentzian metric $g$ that has the components $\eta_{\mu\nu} = \text{diag}\,[+, -, -, -]$ in an orthonormal frame ([1]). However, since we are not immediately concerned with topological issues, we shall use four-dimensional Minkowski space $\mathfrak{M}^4 = (\mathbb{R}^4, \eta)$ for the base manifold $M$. Hence, the total space $P$ can be treated as the product manifold $\mathbb{R}^4 \times U(1)$.

When the Lie group $U(1)$ is regarded as the manifold $S^1$ (viz., a circle), a coordinate for a point on that circle will be described by an angle $\phi$, suitably-defined. Thus, a local coordinate chart on $P$ will look like $(x^\mu, \phi)$. The projection $\pi: \mathbb{R}^4 \times U(1) \to \mathbb{R}^4$, $(x^\mu, \phi) \mapsto x^\mu$ will merely drop the gauge coordinate $\phi$ then.

One will then have a natural frame field $\partial_A = \{\partial_\mu, \partial_\phi\}$ and a natural coframe field $dx^A = \{dx^\mu, d\phi\}$ on $\mathbb{R}^4 \times U(1)$ that correspond to $\{\partial_\mu\}$ and $\{dx^\mu\}$ on $\mathbb{R}^4$. Hence, the considerations that were presented above for $\mathbb{R}^{n+1}$ can be applied to $P = \mathbb{R}^4 \times S^1$ now.

*a. The basic pseudo-hypersurface.* – The basic 1-form that gives the Pfaff equation for our pseudo-hypersurface is the electromagnetic potential 1-form $\bar{A}$, which is initially defined on $P$ to take the form:

---

([1]) Of course, that makes the extension of this theory to the "pre-metric" formulation of electromagnetism [**35**] an important problem to pose.

$$\bar{A} = A + A_\phi \, d\phi, \qquad A = A_\mu(x, \phi) \, dx^\mu, \qquad A_\phi = A_\phi(x, \phi). \tag{5.1}$$

However, a basic property of a connection 1-form on a $U(1)$-principal bundle is that it must be "equivariant" under the action of $U(1)$ on the fibers, which is just $U(1)$ in the present case. If the circle $S^1$ that represents $U(1)$ is the unit circle in the complex plane then the angle $\phi$ will correspond to the point $e^{i\phi}$. The multiplication of planar rotations is then defined by the multiplication of complex numbers: $e^{i\phi} e^{i\phi'} = e^{i(\phi+\phi')}$. Hence, one can just as well represent the composition of the planar rotations by the addition of the angles $\phi + \phi'$. The effect of the equivariance of $\bar{A}$ under that group action is to say that $A_\mu$ cannot be a function of $\phi$ and $A_\phi$ must be a constant. Hence, $\bar{A}$ will take the form:

$$\bar{A} = A + d\phi, \quad A = A_\mu(x) \, dx^\mu. \tag{5.2}$$

The Pfaff equation $\bar{A} = 0$ then defines a hyperplane $\mathfrak{H}_p$ in each tangent space $T_p P$ in the usual way, namely, if $\mathbf{X} \in T_p P$ then $\mathbf{X} \in \mathfrak{H}_p P$ iff:

$$0 = \bar{A}(\mathbf{X}) = A_\mu X^\mu + X^\phi. \tag{5.3}$$

Hence, our basic pseudo-hypersurface will be represented by the horizontal sub-bundle $\mathfrak{H}(P) \to P$ of the tangent bundle $T(P)$. There is then a corresponding decomposition of $T(P)$ into a direct sum $\mathfrak{H}(P) \oplus V(P)$, where the "vertical" sub-bundle $V(P)$ is defined intrinsically by the projection $\pi: P \to M$. Namely, if we differentiate that map at each $p \in P$ then one will get a linear map $d\pi|_p : T_p(P) \to T_{\pi(p)} M$ that takes any vertical tangent vector to zero. In the present simplified case, a vertical vector will take the form:

$$\mathbf{V} = X^\phi \partial_\phi. \tag{5.4}$$

When one has defined a complementary sub-bundle to $V(P)$ in the form of $\mathfrak{H}(P)$, $d\pi|_p$ will represent a linear isomorphism of $\mathfrak{H}_p(P)$ with $T_{\pi(p)} M$. Hence, the tangent vector $\mathbf{X} = X^\mu \partial_\mu + X^\phi \partial_\phi$ will project to $X^\mu \partial_\mu$ under $d\pi|_p$.

In effect, the horizontal subspaces of $T(P)$ look like the tangent spaces to $\mathbb{R}^4$, except that each point of $\mathbb{R}^4$ has one tangent space and a single infinitude of horizontal hyperplanes. Furthermore, there is another important difference between $T(\mathbb{R}^4)$ and $\mathfrak{H}$, namely, that a vector $\mathbf{X} = X^\mu \partial_\mu + X^\phi \partial_\phi$ in $T_p(P)$ will belong to $T_p(\mathbb{R}^4) \subset T_p(P)$ iff $X_\phi = 0$, while, from (5.3), it will belong to $\mathfrak{H}_p$ iff $X_\phi = -A_\mu X^\mu$. Therefore, in some sense, the hyperplane $\mathfrak{H}_p$ is a "tilted" version of $T_{\pi(p)}(\mathbb{R}^4)$.

The degree of integrability of the Pfaff equation $\bar{A} = 0$ is then determined by the first vanishing form in the sequence $\bar{F} \equiv d_\wedge \bar{A}$, $\bar{A} \wedge \bar{F}$, $\bar{F} \wedge \bar{F}$, $\bar{A} \wedge \bar{F} \wedge \bar{F}$, which terminates with a 5-form on $P$, since $P$ is five-dimensional. The 2-form $\bar{F}$ is defined by the exterior derivative of $\bar{A}$:

$$\bar{F} = \bar{d}_\wedge \bar{A}, \tag{5.5}$$

in which the overbar on $d$ indicates that we are using the five-dimensional differential, not the four-dimensional one. Therefore, since:

$$\bar{d}_\wedge \bar{A} = d_\wedge A + \bar{d}_\wedge \bar{d}\phi = d_\wedge A,$$

we see that:

$$\bar{F} = d_\wedge A = \tfrac{1}{2}(\partial_\mu A_\nu - \partial_\nu A_\mu)\, dx^\mu\, dx^\nu, \tag{5.6}$$

and the 2-form $\bar{F}$ will be the same for any choice of the gauge coordinate $\phi$.

If $\bar{F}$ vanishes then $\bar{A}$ will be an exact 1-form on $P$ ([1]). Hence, there will be a smooth function $\psi$ on $P$ such that:

$$\bar{A} = \bar{d}\psi = \partial_\mu \psi\, dx^\mu + \partial_\phi \psi\, d\phi, \tag{5.7}$$

which will then demand that one must have $\partial_\phi \psi = 1$. In that case, the pseudo-hypersurface that $\bar{A}$ defines will be any of the hypersurfaces that are defined by setting $\psi$ equal to a constant, so the degree of integrability will be four.

If $\bar{F}$ does not vanish then the next test of integrability is whether:

$$\bar{A} \wedge \bar{d}_\wedge \bar{A} = \bar{A} \wedge \bar{F} = A \wedge F + d\phi \wedge F \tag{5.8}$$

vanishes. However, since the two 3-forms in this sum are linearly-independent, they would both have to vanish:

$$A \wedge F = 0, \qquad d\phi \wedge F = 0. \tag{5.9}$$

Since the 2-form $F$ does not have $d\phi$ as a factor, and $d\phi$ does not vanish, in general, the only way that $d\phi \wedge F$ could vanish is if $F$ vanished. Hence, we conclude that when $F$ is not identically zero, the pseudo-hypersurface that is defined by $\bar{A} = 0$ will not be completely integrable.

The next test of integrability is whether:

$$\bar{d}_\wedge \bar{A} \wedge \bar{d}_\wedge \bar{A} = F \wedge F \tag{5.10}$$

---

([1]) Strictly speaking, since the manifold $\mathbb{R}^4 \times S^1$ is homotopically equivalent to $S^1$, which is not simply connected, the only thing that one can infer from the vanishing of $\bar{F}$ is that $\bar{A}$ must be *closed*. However, if one removes just one point from the circle (say, $\phi = \pi$), it will become simply connected, and closed 1-forms will all be exact in that case.

vanishes. In such a case, $F$ will be decomposable, so there will be two non-collinear 1-forms $\alpha$ and $\beta$ such that

$$F = \alpha \wedge \beta. \tag{5.11}$$

Many of the most important electromagnetic fields have field strength 2-forms of this type, such as electrostatic fields, magnetostatic fields, and the fields of electromagnetic waves. Hence, the vanishing of $F \wedge F$ is a physically-meaningful condition to impose on $F$. When that condition is satisfied, $\bar{A}$ will take the form:

$$\bar{A} = d\phi + \mu \, d\nu, \tag{5.12}$$

and the integral submanifolds of $\bar{A} = 0$ will be three-dimensional, not four. Namely, they will be the intersections of the level hypersurfaces of the functions $\phi$ and $\nu$.

The next possibility for the integrability of that Pfaff equation is that $F \wedge F$ is non-vanishing, but:

$$\bar{A} \wedge F \wedge F = A \wedge F \wedge F + d\phi \wedge F \wedge F = d\phi \wedge F \wedge F = 0.$$

(The term $A \wedge F \wedge F$ must vanish, since it is a 5-form on a four-dimensional vector space.) Since $F \wedge F$ does not contain $d\phi$ as a factor and $d\phi$ is non-vanishing, in general, this condition must revert to the previous one that $F \wedge F$ must vanish.

Finally, $\bar{A} \wedge F \wedge F = d\phi \wedge F \wedge F$ might be non-vanishing, while $F \wedge F \wedge F$ will have to vanish due to dimensional reasons. In this case, $F \wedge F$ will be non-vanishing, so $F$ cannot be decomposable; i.e., it cannot take the form (5.11), but must take the form:

$$F = \alpha_1 \wedge \beta_1 + \alpha_2 \wedge \beta_2, \tag{5.13}$$

in which the 1-forms $\alpha_1, \beta_1, \alpha_2, \beta_2$ are all linearly independent. The corresponding form of $\bar{A}$ will be:

$$\bar{A} = d\phi + \mu_1 \, d\nu_1 + \mu_2 \, d\nu_2. \tag{5.14}$$

The integral submanifolds that such an $\bar{A}$ defines will then be two-dimensional. Namely, they will be the intersections of the level hypersurfaces of the functions $\phi, \nu_1, \nu_2$.

Typically physics works with the versions of these forms above that one obtains by choosing a section $s : \mathbb{R}^4 \to , x^\mu \mapsto \mathbb{R}^4 \times S^1, x^\mu \mapsto (x^\mu, \phi(x))$ and pulling $\bar{A}$ down to a 1-form on $\mathbb{R}^4$ by way of $s$:

$$s^*\bar{A} = A_\mu(x) \, dx^\mu + d\phi = (A_\mu + \partial_\mu \phi) \, dx^\mu. \tag{5.15}$$

We shall also represent $A_\mu(x) \, dx^\mu$ by $A$, even though the components are defined on points of space-time, not points of $P$, now. Similarly, $\phi$ will now be a function of $x$.

One pulls $\bar{F}$ down to a 2-form on $\mathbb{R}^4$:

$$F = s^*\bar{F} = d_\wedge A = \tfrac{1}{2}(\partial_\mu A_\nu - \partial_\nu A_\mu) \, dx^\mu \wedge dx^\nu \tag{5.16}$$

that is commonly identified as the electromagnetic field strength 2-form.

In order for $s$ to represent an integral submanifold of $\bar{A} = 0$, one would need to have the vanishing of $s^*\bar{A} = A + d\phi$. Hence, $A$ would have to take the form:

$$A = -d\phi, \tag{5.17}$$

which would make:

$$F = 0. \tag{5.18}$$

Such a field $F$ would be referred to as a "pure gauge" field, so generally the main obstruction to the integrability of the Pfaff equation $\bar{A} = 0$ is the existence of a non-vanishing electromagnetic field.

*b. The two fundamental forms on $\mathfrak{H}$.* – Typically, the first fundamental form on $\mathfrak{H}$ (i.e., the metric) is defined first, and the actual metric on $P$ is not as relevant to physics. That is because when one has a metric $g$ on $\mathbb{R}^4$ (i.e., on its tangent bundle) and a smooth projection $\pi : \mathbb{R}^4 \times S^1 \to \mathbb{R}^4$, one can pull $g$ up to $\mathbb{R}^4 \times S^1$ using $\pi$ to get a symmetric doubly-covariant tensor field $\pi^*g$ on $\mathbb{R}^4 \times S^1$. Although it will not be non-degenerate, nonetheless, its restriction to $\mathfrak{H}$ will be non-degenerate, and one can again represent it in local coordinates in the form:

$$\pi^*g\,(p) = g_{\mu\nu}(x)\,dx^\mu\,dx^\nu, \qquad x = \pi(p). \tag{5.19}$$

Since the (vertical) complement $V(P)$ to the horizontal sub-bundle $\mathfrak{H}$ has one-dimensional fibers, the only metric that one can give it must be Euclidian, and we then extend the metric $\pi^*g$ on $\mathfrak{H}$ to a metric $\bar{g}$ on $T(P)$ by simply adding that Euclidian metric to the latter:

$$\bar{g} = \pi^*g + \delta = g_{\mu\nu}\,dx^\mu\,dx^\nu + d\phi\,d\phi. \tag{5.20}$$

In order to get the second fundamental form for our hypersurface $\mathfrak{H}$, we first polarize the differential $d\bar{A}$:

$$d\bar{A} = dA = H + F, \tag{5.21}$$

so $H$ looks like:

$$H = \tfrac{1}{2}(\partial_\mu A_\nu + \partial_\nu A_\mu)\,dx^\mu\,dx^\nu, \qquad A_\mu = A_\mu(x), \tag{5.22}$$

in a coordinate chart on $P$, and $F$ is as above.

When one has a section $s : \mathbb{R}^4 \to P$, one can pull $d\bar{A}$ down to a doubly-covariant tensor field $s^*d\bar{A} = dA$ on $\mathbb{R}^4$ for which $H$ and $F$ take the same forms that they do on $P$. Hence, all that has changed is that the points $p$ in $P$ have been replaced with the points $x = \pi(p)$ in $M$. Of course, here we are getting into uncharted territory as far as physics is concerned, since the 1-form $A$ is always treated as something that is too ambiguous to be worthy of consideration in its own right, so although one might think of $H$ as something analogous to an infinitesimal strain tensor that one would associate with $A$, and proceed with the business of finding its eigenvalues and eigenvectors – i.e., the principal curvatures of $\mathfrak{H}$ and their directions – nonetheless, one would also have to be skeptical about the physical interpretation of those quantities.

*c. Curves in $\mathfrak{H}$.* – If $\tau'$ is any curve parameter then a curve $x(\tau')$ in $P$ is said to be *horizontal* iff its velocity vector lies in the horizontal hyperplane at each of its points. Hence, one will always have:

$$0 = \bar{A}(\dot{\bar{\mathbf{x}}}) = A_\mu \frac{dx^\mu}{d\tau'} + \frac{d\phi}{d\tau'} \tag{5.23}$$

or

$$\frac{d\phi}{d\tau'} = -A(\dot{\mathbf{x}}) \tag{5.24}$$

as a constraint on its velocity $\dot{\bar{\mathbf{x}}}$. That will imply the corresponding constraint on the acceleration:

$$\bar{A}(\ddot{\bar{\mathbf{x}}}) = -\bar{H}(\dot{\bar{\mathbf{x}}}, \dot{\bar{\mathbf{x}}}) = -H(\dot{\mathbf{x}}, \dot{\mathbf{x}}). \tag{5.25}$$

As was pointed out above, although one can speak of the lines of curvature of $\mathfrak{H}$, which then correspond to the principal directions of $H$ at each point, one has a dearth of pre-existing physical examples to serve as a basis for interpreting such curves, and similarly for the asymptotic curves of $\mathfrak{H}$.

If we wish to examine the normal curves on $\mathfrak{H}$ then we first proceed naively by simply applying the definitions and basic equations that we have been using all along. Thus, in order to define normal curves in $\mathfrak{H}$, we must first define the normal vector field $\bar{\mathbf{A}}$ that is metric-dual to the constraint 1-form $\bar{A}$ using the metric $\bar{g}$ on $P$. Hence, the components of $\bar{\mathbf{A}}$ with respect to a natural frame field on a coordinate chart in $P$ will be:

$$A^\mu = g^{\mu\nu} A_\nu, \qquad \mu, \nu = 0, 1, 2, 3, \qquad A^\phi = A_\phi = 1. \tag{5.26}$$

Hence, a vector $\bar{\mathbf{X}} = \mathbf{X} + \mathbf{X}_\phi$ in $T_pP$ will be orthogonal to $\bar{\mathbf{A}} = \mathbf{A} + \partial_\phi$ iff:

$$0 = \bar{g}(\bar{\mathbf{A}}, \bar{\mathbf{X}}) = g(\mathbf{A}, \mathbf{X}) + X_\phi = A(\mathbf{X}) + X_\phi = \bar{A}(\bar{\mathbf{X}});$$

i.e., $\bar{\mathbf{X}} \in \mathfrak{H}$.

If one normalizes $\bar{\mathbf{A}}$ to a unit vector $\mathbf{N}$ in the usual way then the definition of a normal curve, namely, $\ddot{\mathbf{x}} = \|\ddot{\mathbf{x}}\| \mathbf{N}$ will become essentially:

$$\ddot{\mathbf{x}} = \frac{\|\ddot{\mathbf{x}}\|}{\|\bar{\mathbf{A}}\|} \bar{\mathbf{A}};$$

i.e.:

$$\ddot{x}^\mu = \frac{\|\ddot{\mathbf{x}}\|}{\|\overline{\mathbf{A}}\|} A^\mu, \qquad \ddot{x}^\phi = \frac{\|\ddot{\mathbf{x}}\|}{\|\overline{\mathbf{A}}\|},$$

but the second one can be substituted in the first one to get:

$$\ddot{x}^\mu = \ddot{x}_\phi A^\mu. \tag{5.27}$$

Of course, that equation seems to have a rather unphysical character, since it essentially makes acceleration proportional to a potential, when one typically expects it to be coupled to the exterior derivative of that potential. However, as we shall now see, the constrained geodesic equation that is defined by our pseudo-hypersurface does have a fundamental physical significance.

*d. Geodesics in $\mathfrak{H}$.* – One advantage of going to the relativistic formulation of electromagnetism is that it allows one to be simultaneously dealing with (proper) time as a curve parameter and still assume constant-speed. That is because when a curve $x : \mathbb{R} \to \mathfrak{M}^4$, $\tau \mapsto x(\tau)$ is parameterized by proper-time $t$, its velocity vector $\dot{\mathbf{x}}(\tau)$ must always satisfy:

$$c^2 = v^2 = g(\dot{\mathbf{x}}, \dot{\mathbf{x}}) = g_{\mu\nu} \frac{dx^\mu}{d\tau} \frac{dx^\nu}{d\tau}. \tag{5.28}$$

Of course, since we are presently looking at curves in $\mathfrak{H}$, not curves in Minkowski space, we have to decide what to do with the fifth component of $\dot{\overline{\mathbf{x}}}$, namely, $d\phi / d\tau = - A(\dot{\mathbf{x}})$.

Based upon a largely-heuristic argument, we assume that the speed of the curve is gauge invariant, so $v = v(\dot{\mathbf{x}})$, and then form the Lagrangian for a geodesic that is constrained to $\mathfrak{H}$ by way of:

$$\mathcal{L}(x, \phi, \dot{\mathbf{x}}, \dot{\phi}) = v(\dot{\mathbf{x}}) + \lambda(x, \phi)[A(\dot{\mathbf{x}}) + \dot{\phi}] = \mathcal{L}_v(x, \phi, \dot{\mathbf{x}}) + \mathcal{L}_\phi(x, \phi, \dot{\phi}), \tag{5.29}$$

in which the dot means a derivative with respect to an arbitrary curve parameter.

We then get:

$$\frac{\delta v}{\delta x^\mu} = \frac{1}{v}(\ddot{x}_\mu - \dot{v} t_\mu). \qquad \frac{\delta v}{\delta \phi} = 0. \tag{5.30}$$

We also have:

$$\frac{\delta}{\delta x^\mu}[\lambda A(\dot{\mathbf{x}})] = \frac{\delta}{\delta x^\mu}[\lambda A_\mu \dot{x}^\mu] = \frac{\overline{d}}{d\tau}(\lambda A_\mu) - \partial_\mu(\lambda A_\nu) \dot{x}^\nu$$

$$= \frac{\overline{d}\lambda}{d\tau} A_\mu + \lambda \frac{\overline{dA_\mu}}{d\tau} - (\partial_\mu \lambda) A_\nu \dot{x}^\nu - \lambda(\partial_\mu A_\nu) \dot{x}^\nu$$

$$= \frac{\overline{d}\lambda}{d\tau} A_\mu + \lambda(\partial_\mu A_\nu - \partial_\mu A_\nu) \dot{x}^\nu = \frac{\overline{d}\lambda}{d\tau} A_\mu + \lambda F_{\mu\nu} \dot{x}^\nu,$$

so the vanishing of $\delta \mathcal{L}_v / \delta x^\mu$ will give:

$$\frac{1}{c}\ddot{x}_\mu = \frac{\bar{d}\lambda}{d\tau}A_\mu + \lambda F_{\mu\nu}\dot{x}^\nu, \tag{5.31}$$

when we revert to proper-time parameterization.

Meanwhile

$$\frac{\delta \mathcal{L}_\phi}{\delta \phi} = \frac{\delta}{\delta \phi}\left[\lambda \frac{d\phi}{d\tau}\right] = \frac{\bar{d}\lambda}{d\tau} - \frac{\partial \lambda}{\partial \phi}\frac{d\phi}{d\tau} = \frac{\partial \lambda}{\partial x^\mu}\frac{dx^\mu}{d\tau} = \frac{d\lambda}{d\tau},$$

so the vanishing of $\delta \mathcal{L}_\phi / \delta x$ will give:

$$\frac{d\lambda}{d\tau} = 0, \tag{5.32}$$

which is basically a conservation law for $\lambda$; namely, it must be constant along the integral curves. Hence, the equation of a geodesic that is constrained to $\mathfrak{H}$ will take the form:

$$\ddot{x} = (\dot{\phi}\partial_\phi \lambda)A - c\lambda i_{\dot{x}}F, \tag{5.33}$$

which basically agrees with (4.19).

The acceleration $\ddot{x}$ will be normal iff $A \wedge \ddot{x}$ vanishes. However, when one takes the exterior product of $A$ with the right-hand side of (5.33), that will give $-c\lambda A \wedge i_{\dot{x}}F = c\lambda i_{\dot{x}}(A \wedge F)$, so $\ddot{x}$ will be normal iff $i_{\dot{x}}(A \wedge F)$ vanishes. Thus, when the basic pseudo-hypersurface is not completely integrable, it is possible for geodesics to not be normal curves.

When the Lagrange multiplier $\lambda$ is chosen to be the constant $-q/mc$, equation (5.33) will take the form:

$$m\ddot{x} = q i_{\dot{x}}F, \tag{5.34}$$

which is precisely that of the Lorentz force equation for a point-particle of charge $q$ and mass $m$.

One might argue that typically Lagrange multipliers are not constants, but one simple solution to that dilemma is to remember that point-like charges can be regarded as approximations to extended charges. Hence, instead of setting $\lambda(x)$ equal to the constant $q/m$, one might replace $q$ with a charge density $\sigma(x)$ and $m$ with a mass density $\rho(x)$. Of course, in order for their quotient to still be a function of $x$, one cannot assume the approximation that $\sigma(x) = q\,n(x)$ and $\rho(x) = m\,n(x)$, where $n(x)$ is a common density function that describes how both are distributed over time and space.

When $\lambda = \sigma/\rho$, the condition (5.32) will take the form

$$0 = \frac{d\lambda}{d\tau} = \frac{1}{\rho}\left(\frac{d\sigma}{d\tau} - \frac{\sigma}{\rho}\frac{d\rho}{d\tau}\right). \tag{5.35}$$

At any $x$ where $\rho(x)$ does not vanish, (5.35) can be satisfied by the pair of continuity equations for the charge and mass densities:

$$0 = \frac{d\sigma}{d\tau} = \frac{d\rho}{d\tau}, \tag{5.36}$$

which are entirely reasonable from a physical standpoint.

In order to eliminate the normal term from the right-hand side of (5.33), one might make the equally-plausible assumption that $\lambda$ (or $\rho$ and $\sigma$) is independent of $\phi$. Indeed, conservation of charge is usually regarded as a consequence of gauge-invariance, at least in the eyes of Noether's theorem.

When (5.35) is satisfied, (5.33) will reduce to the form:

$$\rho \ddot{x} = \sigma i_{\dot{x}} F. \tag{5.37}$$

Perhaps the real innovation that comes from interpreting the trajectories of charges moving in an electromagnetic field as geodesics for non-holonomic constraints is that it gives one a deeper insight into the term $A(\mathbf{J})$ in the Lagrangian that basically describes the coupling of the charge-current $\mathbf{J}$ to the electromagnetic field that it lives in. In effect, we can now see that coupling as a coupling of the motion of the charge to a non-holonomic constraint that is defined by that external field.

**6. Discussion.** – From what was developed in this study, it is clear that the scope of the applications of pseudo-hypersurfaces to physics is essentially identical to the scope of application of the theory of Pfaff equation to physics, which the author has previously discussed [**25**]. Hence, there are still many other physical applications of pseudo-hypersurfaces to explore.

There are some limitations to the concept of a pseudo-hypersurface that are based upon limitations to the theory of the Pfaff equation itself. One possible direction of extension of the theory is from a single Pfaff equation to a system of Pfaff equations. One might call the sub-bundle of the tangent bundle to a manifold that is defined by a system of Pfaff equations a "pseudo-submanifold" accordingly.

Indeed, constrained motion, especially non-holonomically constrained motion, is usually treated as something that involves more than one constraint, and therefore more than one Pfaff equation. For instance, the "canonical" example of a mechanical system with non-holonomic constraint, namely, a wheel rolling without slipping on a plane or surface, actually requires two Pfaff equations in order to define the constraint. Although Frobenius's theorem for complete integrability can be extended to systems of Pfaff equations, the determination of the degree of integrability is not as straightforward in the general case. However, one will still have integral curves in pseudo-submanifolds, even when those pseudo-submanifolds are not completely integrable.

Another obvious direction of expansion for the theory and applications of pseudo-hypersurfaces is from the linear case that we have treated to the nonlinear case. In the way that Lie [**36**] envisioned these matters, the Pfaff equation is a special case of the Monge equation, which amounts to the vanishing of a function that is defined on the cotangent bundle and is homogeneous of some degree on its fibers. The specialization to the Pfaff equation is obtained by specifying that the function should be linear on the fibers. The nonlinear examples in physics include dispersion laws for the propagation of waves, such as electromagnetic waves, and Hamiltonian functions.

Clearly, the last example opens up a vast number of physical applications, in its own right. At any rate, it should be clear that the concept of a pseudo-hypersurface plays a very fundamental role in the mathematical models for many physical phenomena.

## References (*)

---

(*) Many of the non-English references are available in English translations at the author's website: neo-classical-physics.info, which is updated on an ongoing basis.

__________